\documentclass{aa}  
\usepackage{graphicx}
\usepackage{txfonts,textcomp}
\usepackage{hyperref}
\newbox\grsign \setbox\grsign=\hbox{$>$} \newdimen\grdimen \grdimen=\ht\grsign
\newbox\simlessbox \newbox\simgreatbox
\setbox\simgreatbox=\hbox{\raise.5ex\hbox{$>$}\llap
     {\lower.5ex\hbox{$\sim$}}}\ht1=\grdimen\dp1=0pt
\setbox\simlessbox=\hbox{\raise.5ex\hbox{$<$}\llap
     {\lower.5ex\hbox{$\sim$}}}\ht2=\grdimen\dp2=0pt

\def\simless{\mathrel{\copy\simlessbox}}
\newbox\simppropto
\setbox\simppropto=\hbox{\raise.5ex\hbox{$\sim$}\llap
     {\lower.5ex\hbox{$\propto$}}}\ht2=\grdimen\dp2=0pt

\begin{document}

\title{\fontsize{14.3}{10}\selectfont{Abundances of iron-peak elements in 58 bulge spheroid stars from APOGEE}}
\titlerunning{Abundances of iron-peak elements in 58 bulge spheroid stars from APOGEE}
 \authorrunning{B. Barbuy et al.}
 
 \author{\fontsize{10}{10}\selectfont{
 B. Barbuy \inst{1} 
 \and A.C.S. Fria\c ca\inst{1} 
 \and H. Ernandes \inst{2}
  \and P. da Silva \inst{1} 
 \and S. O. Souza \inst{3} 
 \and  J. G. Fern\'andez-Trincado \inst{4}  
\and K. Cunha\inst{5,6} 
\and V. V. Smith\inst{7}
\and T. Masseron\inst{8,9}
\and A. P\'erez-Villegas\inst{10}
\and C. Chiappini\inst{11}
\and A.B.A. Queiroz\inst{8}
\and B. X. Santiago\inst{12}
\and T. C. Beers\inst{13}
\and F. Anders\inst{14,15,16}
\and R. P. Schiavon\inst{17}
\and M. Valentini\inst{11}
\and D. Minniti\inst{18,19}
\and D. Geisler\inst{20,21,22}
\and D. Souto\inst{23}
\and V. M. Placco\inst{7}
\and M. Zoccali\inst{24}
\and S. Feltzing\inst{2}
\and M. Schultheis\inst{25}
\and C. Nitschelm\inst{26}
}
}

\institute{
Universidade de S\~ao Paulo,  IAG, Departamento de Astronomia, 05508-090 S\~ao Paulo, Brazil
\and 
Lund Observatory, Department of Astronomy and Theoretical Physics, Lund University, Box 43, SE-221 00 Lund, Sweden 
\and 
Max Planck Institute for Astronomy, K\"onigstuhl 17, D-69117 Heidelberg, Germany
\and
Instituto de Astronom\'ia, Universidad Cat\'olica del Norte, Av. Angamos 0610, Antofagasta, Chile
\and
 University of Arizona, Steward Observatory, Tucson, AZ 85719, USA 
 \and
Observat\'orio Nacional,  rua General Jos\'e  Cristino 77, S\~ao Crist\'ov\~ao, Rio de Janeiro 20921-400, Brazil
\and
NSF NOIRLab, 950 N. Cherry Ave., Tucson, AZ 85719, USA 
\and
 Instituto de Astrof\'{\i}sica de Canarias, C/Via Lactea s/n, E-38205 La Laguna, Tenerife, Spain 
\and
Departamento de Astrof\'{\i}sica, Universidad de La Laguna, E-38206 La Laguna, Tenerife, Spain
\and
Instituto de Astronom\'ia, Universidad Nacional Aut\'onoma de M\'exico, A. P. 106, C.P. 22800, Ensenada, B. C., M\'exico
\and
Astrophysikalisches Institut Potsdam, An der Sternwarte 16, Potsdam, 14482, Germany
\and
Universidade Federal do Rio Grande do Sul, Caixa Postal 15051, 91501-970 Porto Alegre, Brazil
\and
Department of Physics and Astronomy and JINA Center for the Evolution of the Elements (JINA-CEE), University of Notre Dame, Notre Dame, IN 46556  USA
\and
Departament de F\'{\i}sica Qu\~antica i Astrof\'{\i}sica (FQA), Universitat de Barcelona (UB), Mart\'{\i} i Franqu\`es, 1, 08028 Barcelona, Spain
\and
 Institut de Ci\`encies del Cosmos, Universitat de Barcelona (IEEC-UB), Mart\'{\i} i Franqu\`es 1, 08028 Barcelona, Spain
\and
Institut d'Estudis Espacials de Catalunya (IEEC), Edifici RDIT, Campus UPC, 08860 Castelldefels (Barcelona), Spain
 \and
Astrophysics Research Institute, Liverpool John Moores University, Liverpool, L3 5RF, UK 
\and 
Instituto de Astrof\'isica, Facultad de Ciencias Exactas, Universidad Andres Bello, Fern\'andez Concha 700, Las Condes, Santiago, Chile 
\and 
Vatican Observatory, Vatican City State 00120, Italy 
\and
Departamento de Astronomia, Casilla 160-C, Universidad de Concepcion, Chile 
\and
Instituto de Investigaci\'on Multidisciplinario en Ciencia y Tecnolog\'ia, Universidad de La Serena. Avenida Ra\'ul Bitr\'an S/N, La Serena, Chile 
\and
Departamento de Astronom\'ia, Facultad de Ciencias, Universidad de La Serena. Av. Juan Cisternas 1200, La Serena, Chile 
 \and
 Universidade Federal de Sergipe, Av. Marechal Rondon, S/N, 49000-000
S\~ao Crist\'ov\~ao, SE, Brazil
\and
Instituto de Astrof\'isica, Pontificia Universidad Cat\'olica de Chile, Vicu\~na Mackenna 4860, Macul, Casilla 306, Santiago 22, Chile
 \and
 Universit\'e C\^ote d'Azur, Observatoire de la C\^ote d'Azur, CNRS, Laboratoire Lagrange, Nice, France 
 \and
Centro de Astronom{\'i}a (CITEVA), Universidad de Antofagasta, Avenida Angamos 601, Antofagasta 1270300, Chile 
}
            
   \date{Received ....; accepted .....}

 
  \abstract
 {Stars presently identified in the bulge spheroid are probably very old, and their abundances can be interpreted as due to the fast chemical enrichment of the early Galactic bulge. The abundances of the iron-peak elements are important tracers of nucleosynthesis processes, in particular oxygen burning, silicon burning, the weak $s$-process, and $\alpha$-rich freeze-out.}
   {The aim of this work is to derive the abundances of V, Cr, Mn, Co, Ni, and Cu in 58 bulge spheroid stars and to compare them with the results of a  previous analysis of data from the Apache Point Observatory Galactic Evolution Experiment (APOGEE).}
  {We selected the best lines for V, Cr, Mn, Co, Ni, and Cu located within the \textit{H}-band of the spectrum, identifying the most suitable ones for abundance determination, and discarding severe blends. Using the stellar physical parameters available for our sample from the DR17 release of the APOGEE project, we derived the individual abundances through spectrum synthesis. We then complemented these measurements with similar results from different bulge field and globular cluster stars, in order to define the trends of the individual elements and compare with the results of chemical-evolution models.} 
   {We verify that the \textit{H}-band has useful lines for the
   derivation of the elements V, Cr, Mn, Co, Ni, and Cu in moderately
   metal-poor stars. The abundances, plotted together with others from high-resolution spectroscopy of bulge stars, indicate that: V, Cr, and Ni vary in lockstep with Fe; Co tends to vary in lockstep with Fe, but could be showing a slight decrease with decreasing metallicity; and Mn and Cu decrease with decreasing metallicity. These behaviours are well reproduced by chemical-evolution models that adopt literature yields, except for Cu, which appears to drop faster than the models predict for [Fe/H]$<-$0.8. Finally, abundance indicators combined with kinematical and dynamical criteria appear to show that our 58 sample stars are likely to have originated in situ.}
  {}
   \maketitle

\section{Introduction}

The very early Galactic bulge formed from a merger of dark matter haloes
and their respective matter content, consisting of first-generation stars
and globular clusters (GCs)\citep[e.g.][]{gao10}.
The subsequent processes of bulge formation are still debated in the literature,
and observational evidence is needed to advance discussions on this topic.
The  GCs identified with a metallicity of [Fe/H]$\sim-$1 appear to be
very old, with ages of 12.5 -- 13.5 Gyr \citep{bica24}, and should have formed
in the earliest bulge. In this work, we analyse stars that could be the counterparts
of these clusters.

The Galactic bulge is known to host two main  populations of  field stars: a metal-poor,
alpha-enhanced population ([Fe/H]$\sim-$0.5; [$\alpha$/Fe]$\sim$+0.25), and a metal-rich, alpha-poor
one ([Fe/H]$\sim$+0.3; [$\alpha$/Fe]$\sim$0.0 \citep{rojas-arriagada19,queiroz20}.
The two components have different spatial distribution and kinematics (e.g. 
\citealt{babusiaux10,zoccali17,queiroz21}). Specifically, the metal-rich component is believed to be associated with the Galactic bar, whose origin can certainly be traced back to the well-known bar formation
caused by the dynamical instabilities of the disc, themselves induced by the spiral arms. The metal-poor component, on the other hand, traces a more spheroidal component whose origin is still debated. It could be an early so-called classical bulge, but
some models can reproduce structures qualitatively compatible with this component 
as a result of the early merger of substructures caused by
the dynamical instability of a disc with a rather large velocity dispersion
 (e.g. \citealt{athanassoula17,debattista17}).

Because the differences between the weak-bar or spheroid resulting from a thick disc and a pressure-supported classical bulge seem to be rather subtle, it is important to study the
metal-poor component of the Milky Way (MW) bulge  in great detail in order to provide multidimensional constraints on the models, and thus be able to distinguish between the two formation scenarios.
To this aim, a few years ago  we began to complete and characterise a sample of 58 moderately metal-poor bulge stars that can be safely associated with the old and 
metal-poor bulge spheroid \citep{razera22}.

In order to better understand the earliest bulge stars, we selected stars with a metallicity of [Fe/H]$<-$0.8, because among the metal-poor stars in the Galactic bulge there is a peak at [Fe/H]$\sim-$1.0 in both bulge GCs \citep{rossi15,bica16,perez20,bica24}
and field stars \citep{lucey21}. This relatively high metallicity for the oldest stars
is due to a fast chemical enrichment \citep{Chiappini11,wise12,barbuy18a,matteucci21}. 

There is also the possibility to have bulge stars with an ex situ origin, brought in via
 early accretion events that include the Gaia-Sausage-Enceladus (GSE --
\citealt{belokurov18,helmi18}), 
and other proposed ones; in particular Kraken \citep{kruijssen19}, Koala \citep{forbes20}, Heracles \citep{horta21}, and Aurora \citep{myeong22,belokurov23}, which are the most important structures  identified in the region.

We applied this selection process to the reduced proper motion (RPM) stars from \citet{queiroz21} as a
starting point, with stars observed by the
Apache Point Observatory Galactic Evolution Experiment (APOGEE project -- \citealt{majewski17}).
To select spheroid stars, we  applied kinematical and dynamical criteria.
These criteria applied to APOGEE stars
resulted in 58 stars, the characteristics of which are reported in \citet{razera22}.

Once the stars were selected, we extracted the H-band spectra from APOGEE and reanalysed them.
In \citet{razera22}, we analysed the abundances of C, N, O, Mg, Si, Ca, and Ce; in \citet{barbuy23}, the same stars were analysed for their Na and Al lines in the \textit{H}-band. \citet{sales-silva24} analysed the neutron-capture elements Nd and Ce,
including some stars in common with the present sample.
Our main interest in the present work is to analyse the iron-peak element
abundances of these moderately metal-poor spheroid bulge stars. 
Generally, the $\alpha$-elements and heavy elements tend to be more commonly studied, 
given their relatively easy interpretation in terms of nucleosynthesis. The iron-peak elements 
that are less studied, however, have potentially powerful implications; 
for example, for interpretations regarding the
origin of stars as in situ or ex situ, as recently shown by \citep{nissen24}.

The iron-peak elements with measurable lines in our sample stars
in the \textit{H}-band are V, Cr, Mn, Co, Ni, and Cu.  The elements V, Cr, Mn, and Co are in the lower iron-peak element group, which includes  elements with 21 $\leq$ Z $\leq$ 27
 (45 $\leq$ A $\leq$ 58). In massive stars, 
 depending on temperatures and densities, these elements are produced in explosive
 oxygen burning and incomplete and complete explosive Si burning;
 for densities typical of core-collapse supernovae (CCSNe), $\alpha$-rich freeze-out takes place
 \citep{woosley95,nomoto13}. The elements Ni and Cu are
 among the upper iron-group
 elements, which include elements with 28 $\leq$ Z $\leq$ 32
 (57 $\leq$ A $\leq$ 72). In massive stars, these elements 
 are mainly produced in two processes, namely neutron capture on
 iron-group nuclei during He burning and later burning stages,
 also called weak-$s$ component \citep{limongi03}, and the $\alpha$-rich freeze-out in the deepest layers. The iron-peak elements are also produced by Type Ia supernovae
\citep{iwamoto99}.

This paper is organised as follows: in Section \ref{sec2}, we describe the sample stars. In Section \ref{sec3}, we describe the present abundance analysis.
In Section \ref{sec4}, we  discuss our results. In Section \ref{sec5}, we present chemical-evolution models that are compared with the data. In Section \ref{sec6} we use our derived abundances to identify our sample stars through in situ---ex situ  origin indicators.
We summarise our conclusions in Section \ref{sec7}.

\section{The sample}\label{sec2}

As explained in \citet{razera22}, our selection is based on the
 RPM sample from \cite{queiroz21}, which, in turn,  is based on the stars observed by APOGEE, combined with {\tt{StarHorse}} distances \citep{santiago16, queiroz18}, and cross-matched with proper motions from the  Gaia Early Data Release 3 \citep{gaia21}. The selection identified 
stars with  a distance to the Galactic centre of d$_{\rm GC}< 4$ kpc,  a maximum height of $| Z|_{\rm max}< 3$ kpc, eccentricity of $> 0.7$, and with orbits not supporting the bar, where the orbits were computed in \citet{queiroz21},
and imposing a metallicity of [Fe/H] $<-0.8$.
This led to a sample of 58 stars with spectra observed and analysed within APOGEE.

APOGEE is part of the Sloan Digital Sky Survey IV
\citep[SDSS-IV/V]{blanton17}. The APOGEE 
spectroscopic programs targeted MW stars at high resolution ($R \sim$ 22,500)  and
high signal-to-noise ratios in the \textit{H}-band  (15140-16940 {\rm \AA}) 
\citep{wilson19}
and included about 7$\times$10$^{5}$ stars, covering both the northern and southern sky. While APOGEE-1 observed
the Galactic bulge at $l>0^{\circ}$ using the 2.5m Sloan Foundation Telescope at the Apache Point Observatory in New Mexico \citep{gunn06}, these observations were complemented with
APOGEE-2 using the 2.5m Ir\'en\'ee du Pont Telescope at the Las Campanas Observatory in Chile \citep{bowen73}. \citet{santana21} and \citet{beaton21} describe the targeting of  APOGEE
for the south and north, respectively.

The analysis of \textit{H}-band spectra in the APOGEE project is carried out through a Nelder-Mead algorithm 
\citep{nelder65},
which simultaneously  fits the stellar parameters ---effective temperature (T$_{\rm eff}$), gravity (log~g), metallicity ([Fe/H]), and microturbulence velocity (v$_{\rm t}$)--- together with the abundances of carbon, nitrogen, and $\alpha$-elements with the APOGEE Stellar Parameter and Chemical
Abundances Pipeline (ASPCAP) \citep{garcia-perez16}, which is based on the FERRE code \citep{allende-prieto06}. 
In the present work, we use APOGEE Data Release 17 - DR17 \citep{abdurro22}.

\section{Calculations}\label{sec3}

We adopted the uncalibrated stellar parameters effective temperature
(T$_{\rm eff}$), gravity (log~g), metallicity ([Fe/H]), and microturbulence velocity (v$_{t}$) from the APOGEE DR17 results; these are  reported here in Table \ref{results}. 
For DR17, the parameters were obtained with new spectral grids constructed using 
the Synspec spectrum synthesis code \citep{hubeny17,hubeny21}.
We note that in 
\citet{dasilva24} we verified the reliability of ASPCAP for deriving
stellar parameters, and concluded that the use of molecular-line
intensities is a powerful method, in particular for the derivation
of effective temperatures.

We computed the abundances of V, Cr, Mn, Co, Ni, and Cu in the \textit{H}-band
 using the code {\tt TURBOSPECTRUM} from \citet{alvarez98} and \citet{plez12}. 
Model atmosphere grids are from \citet{gustafsson08}. 
The solar abundances  of the studied iron-peak elements are from \citet{asplund21}, namely:
A(Cr) = 5.62, A(Mn) = 5.42, A(Co) = 4.94, A(Ni) = 6.20, and A(Cu) = 4.18.


Table \ref{linelist} reports the lines in the \textit{H}-band that we used to measure the abundances
of the iron-peak elements V, Cr, Mn, Co, Ni, and Cu in the spectra of the sample stars. 
Oscillator strengths were adopted from the line list of the APOGEE collaboration,
which were initially adopted from  the most recent line
list of \citet{kurucz95}, 
with log gf values updated with NIST values, and  adjusted based on the Sun and
Arcturus but only within 2 sigma of log gf uncertainties. 
For comparison purposes, we also show the log~gf values 
from the Vienna Atomic Line Database (VALD3): see the  line lists of \citet{ryabchikova15} 
and \citet{kurucz95}. We note that NIST log~gf values are not available
for any of these studied lines. The APOGEE lines that are given as split
in hyperfine structure (hfs) are only indicated as such.
The lines identified as found in the present work correspond to a search for
measurable lines using the APOGEE line list; the weak lines for this work could be
useful in spectra of more metal-rich stars.

The  full atomic line list employed is that from the APOGEE collaboration,
together with the molecular lines described in \citet{smith21}. 
Previously, \citet{razera22} and \citet{barbuy23} examined the
 lines of the elements C, N, O, Na, Al, Mg, Si, Ca, and Ti (Ti was considered as unmeasurable in these stars due to severe blends).

\begin{table*}
\centering
\caption[4]{Line list and oscillator strengths. }
\resizebox{0.7\textwidth}{!}{
\begin{tabular}{lccrrrcccccccc}
\hline
\noalign{\smallskip}
\hbox{Species} & \hbox{$\lambda$} & \hbox{$\chi_{ex}$}  &\hbox{log~gf}  &\hbox{log~gf}  &\hbox{log~gf} &\hbox{Comments} & Source & \\
& \hbox{(\AA)} &\hbox{(eV)} & \hbox{(VALD3)}   & \hbox{(Kurucz)} & \hbox{(APOGEE)} & \hbox{} & \hbox{}   \\ 
\noalign{\smallskip}
\hline
\noalign{\smallskip}
\hbox{VI}
& 15924.791$^*$ & 2.138 & $-$1.108  & $-$1.177 & hfs & best line & Smith+13 \\
& 15925.595 & 4.888 & $-$3.134  & --- & hfs &  weak & Hayes+22\\
\hline
\noalign{\smallskip}
\hbox{CrI}  
& 15177.217 & 5.950  & $-$1.140  & $-$1.960 & $-$1.140 & weak  & present search\\
& 15177.759 & 5.978 & $-$2.741  & $-$2.192 & $-$2.741 & weak  & present search \\
& 15178.593$^*$ & 3.369 & $-$2.020  & $-$2.542 & $-$1.651 & 2nd best line & present search \\
& 15680.063$^*$ & 4.697 &  0.068  & 0.270  & $-$0.001 & best line & Smith+13 \\
& 15860.214 & 4.697 & $-$0.063 & 0.129  & $-$0.077 & Apogee gap & Smith+13 \\
& 16015.327 & 4.696 & $-$0.141  & $-$0.105 & $-$0.141 & heavily blended & present search \\
\hline
\noalign{\smallskip}
\hbox{MnI}
& 15159.200$^*$ & 4.889 & 0.619 & 0.606 & hfs & ... & Smith+13 \\
& 15217.793$^*$ & 4.889 & 0.520 & 0.507 & hfs & ... & Smith+13 \\
& 15262.702$^*$ & 4.889 & --- & 0.379 &  hfs & ... & Smith+13 \\
\hline
\noalign{\smallskip}
\hbox{CoI} 
& 16757.711$^*$ & 3.409 &$-$0.923 & $-$1.369 & hfs & unique line & Smith+13\\
\hline
\noalign{\smallskip}
\hbox{NiI} 
& 15605.655$^*$ & 5.305 & $-$0.150 & ---      & $-$0.247 & ... &   Smith+13 \\
& 15632.611 & 5.305 & $-$0.042 & $-$0.247 &  0.074   & wing strong line & Smith+13 \\
& 16013.745$^*$ & 5.305 & $-$0.699  & --- & $-$0.297 & ... & present search \\
& 16136.097$^*$ & 5.488 & $-$0.003 & $-$0.165 & $-$0.156 & ... &  present search\\
& 16153.114     & 5.525 & $-$3.048 & ---      & $-$3.048 & wing strong line & present search \\
& 16363.105$^*$ & 5.283 &  0.588   & 0.422    & 0.274    &  ... & present search \\
& 16584.439$^*$ & 5.305 & $-$0.876 & ---   & $-$0.485 & ... & Smith+13 \\
& 16585.380     & 6.035 & $-$4.052 & $-$2.190 & $-$4.052 & weak &  present search\\
& 16589.440$^*$ & 5.469 & $-$0.345 & $-$0.493 & $-$0.533 & ... & Smith+13 \\
& 16673.583$^*$ & 6.034 & 0.221    & 0.678    & 0.306    & ... & Smith+13 \\
& 16815.471$^*$ & 5.305 & $-$0.584 & $-$0.547 & $-$0.501 & ... & Smith+13 \\
& 16818.746$^*$ & 6.039 & 0.473    & ---      &  0.347   & ... & Smith+13 \\
& 16823.121     & 6.256 & $-$1.389  & ---     & $-$1.389 & weak & present search\\
\hline
\noalign{\smallskip}
\hbox{CuI} 
& 16005.735$^*$ & 5.348 & $-$0.205 & $-$0.050 & $-$0.157 & unique line & Smith+13 \\
\noalign{\hrule\vskip 0.1cm} 
\hline                 
\label{linelist}
\end{tabular}}
\begin{minipage}{13cm}
\vspace{0.1cm}
\small Notes: Oscillator strengths from VALD3,  \citet{kurucz95}, 
and the APOGEE collaboration (adopted) are reported. The present search
used the APOGEE line list. The symbol $^*$ indicates
the lines useful for the present moderately metal-poor stars.
\end{minipage}
\end{table*}

\section{Iron-peak elements: V, Cr, Mn, Co, Ni, and Cu}\label{sec4}

We analysed lines of V, Cr, Mn, Co, Ni, and Cu as detailed below. The fits
were all carried out visually, adopting convolutions with FWHM from 0.65 to 0.75 {\rm \AA} in the range 15,000 to 17,000 {\rm \AA}. These full width at half maximum (FWHM) values are compatible with
those based on a directly measured FWHM of $\sim$0.7 {\rm \AA}, with 10 to 20 per cent
variations seen across the wavelength range by \citet{ashok21} and \citet{nidever15}.
Although for most of the lines the resulting abundance is similar to that reported
in DR17, there are cases where the visual inspection is needed because of noise or defects. The results are given in Table \ref{results}.

\begin{figure}
    \centering
    \includegraphics[width=8.5cm]{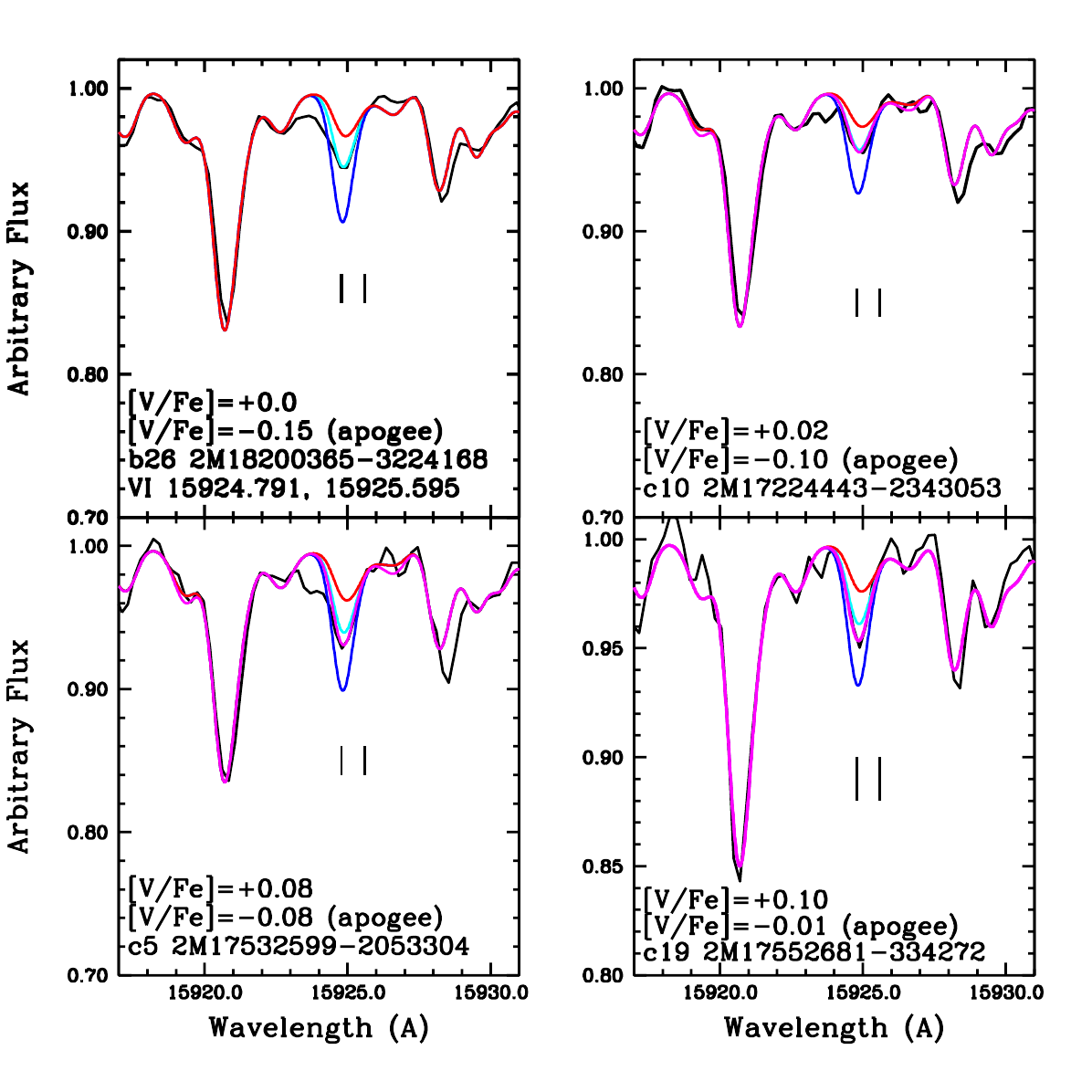}
    \caption{\ion{V}{I} lines in stars 2M18200365-3224168 (b26), 2M17532599-2053304 (c5), 2M17224443-2343053 (c10), and 2M17552681-3342729 (c19)
     computed with  [V/Fe]=$-$0.3 (red), 0.0 (cyan), +0.3 (blue), and
    final values if different (magenta), compared with the observed
    spectrum (black). }
    \label{figv}
\end{figure}

{\it Vanadium:} The \ion{V}{I} 15924.791 {\AA} line, together with the
weaker line at 15925.595 {\rm \AA} is measurable; all other lines are too shallow in these metal-poor stars.
\citet{hayes22} indicate a further two lines, which only appear in 
stars more metal-rich than those analysed here.  Figure 7 of this latter paper shows results for stars
with metallicities [Fe/H]$>-$1.0.
In Table \ref{apogee},
we replace the values of [V/Fe] from the DR17 results with more reliable measurements from the 
BACCHUS Analysis of Weak Lines in APOGEE Spectra (BAWLAS).
BAWLAS is part of the DR17 release as a Value Added Catalog \citep[VAC][]{hayes22}, and
presents abundances for some of the elements represented by only weak lines, namely Na,
P, S, V, Cu, Ce, and Nd, as well as $^{12}$C/$^{13}$C ratios.  The 
VAC used only spectra with higher signal-to-noise ratios (S/N$\sim$120).
Figure \ref{figv} shows the fit to the \ion{V}{I} 15924.791 {\AA} line in four stars:
2M18200365-\-3224168 (b26), 2M17532599-\-2053304 (c5), 2M17224443-\-2343053 (c10), and
2M17552681-\-3342729 (c19). However, even considering that some of the fits are credible, some noise is present in the region
and the plot of [V/Fe] versus [Fe/H] shows a large spread (see Figure \ref{plotv}).

{\it Chromium:} There are four measurable lines, but only two are
useful for all stars. 
This is because,
for one-third of the stars, the line at \ion{Cr}{I} 15860.214 {\rm \AA} falls in the instrument gap and the  line \ion{Cr}{I} 16015.327 {\rm \AA} is blended with
a strong unidentified feature. The latter line is measurable
for only a few stars; in the others, it results in a spurious,
much higher Cr abundance. Regarding the blends disturbing the 16015.327 {\rm \AA}  line listed in Table \ref{linelist}, \citet{smith21} points out possible identifications of
a feature at 16016.75 {\rm \AA}, which could be \ion{Ni}{I}, \ion{Zr}{I}, or \ion{Ni}{I};
it is also blended with a $^{12}$C$^{16}$O line, but the CO lines are taken into account
using the proper abundances of C and O from \citet{razera22}.
 
\begin{figure}
    \centering
    \includegraphics[width=8.5cm]{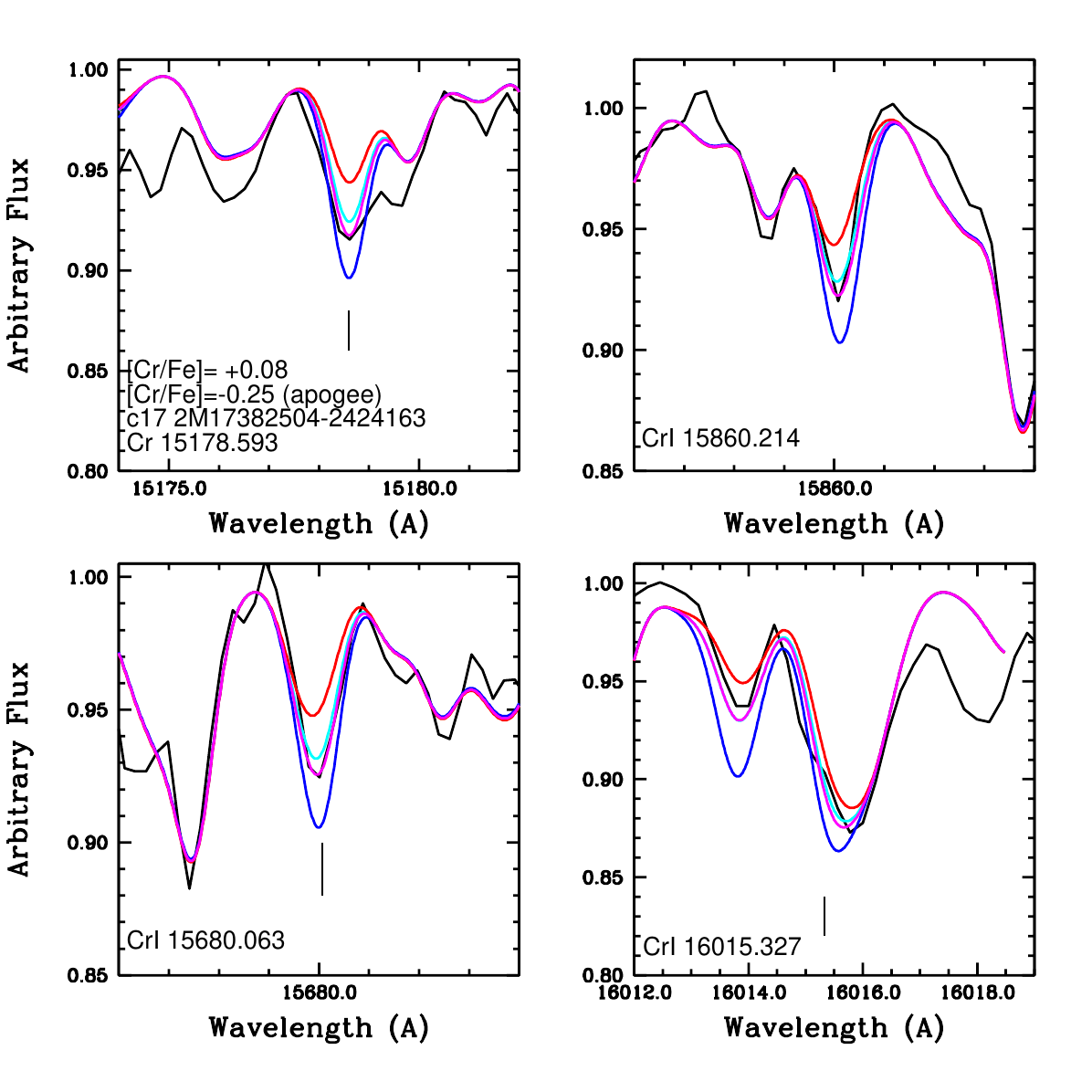}
    \caption{\ion{Cr}{I} lines in star
    2M17382504-2424163 (c17) computed with
    [Cr/Fe]=$-$0.3 (red), 0.0 (cyan), +0.3 (blue), and
    +0.08 (magenta) compared with the observed
    spectrum (black). The final result of [Cr/Fe]=+0.08 is shown in magenta.}
    \label{c17cr}
\end{figure}

We adopted a mean of the
Cr abundance from the two  measurable lines.
With most cases, measured from the \ion{Cr}{I} 15178.593 {\rm \AA}  and 15680.063 {\rm \AA}  lines,
we adopted the mean obtained from the fits to these lines, or only the fit
for the best line, which is \ion{Cr}{I} 15680.063 {\rm \AA}. In some cases,
the result is the mean of the three lines (always discarding \ion{Cr}{I} 16015.327 {\rm \AA,}
except in one unique star).
Figure \ref{c17cr} provides an example of the 
fits to the Cr lines for the star 2M17382504-2424163 (c17).

\begin{figure}
    \centering
    \includegraphics[width=8.5cm]{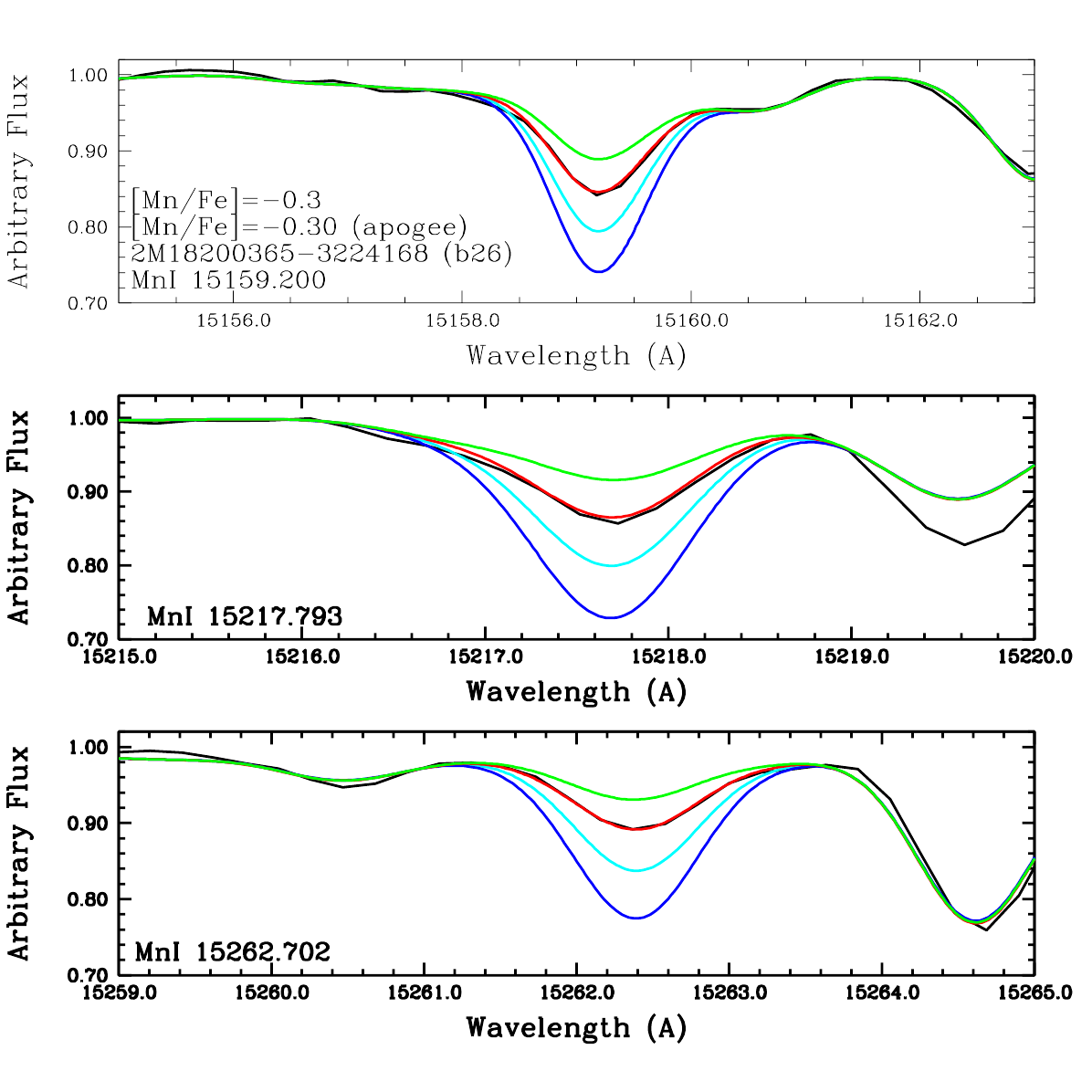}
    \caption{Manganese lines in star 2M18200365-3224168 (b26).
    The synthetic spectra were computed with
    [MnFe]=$-$0.6 (green), $-$0.3 (red), 0.0 (cyan), and +0.3 (blue), and are
    compared with the observed spectrum (black).
    A final value of [MnFe]=-0.3  fits the three lines well. }
    \label{figmn}
\end{figure}

{\it Manganese:} The three lines listed in the APOGEE reference papers cited above
are suitable. Two other lines that we detected are too faint in these
metal-poor stars. In general, we find very good agreement with
the ASPCAP results (see Table \ref{apogee}).
Figure \ref{figmn} shows a fit for the 3 \ion{Mn}{I} lines in star
2M18200365-3224168 (b26). In cases where the Mn abundance varies among
the lines, a mean value was adopted.

\begin{figure}
    \centering
    \includegraphics[width=8.5cm]{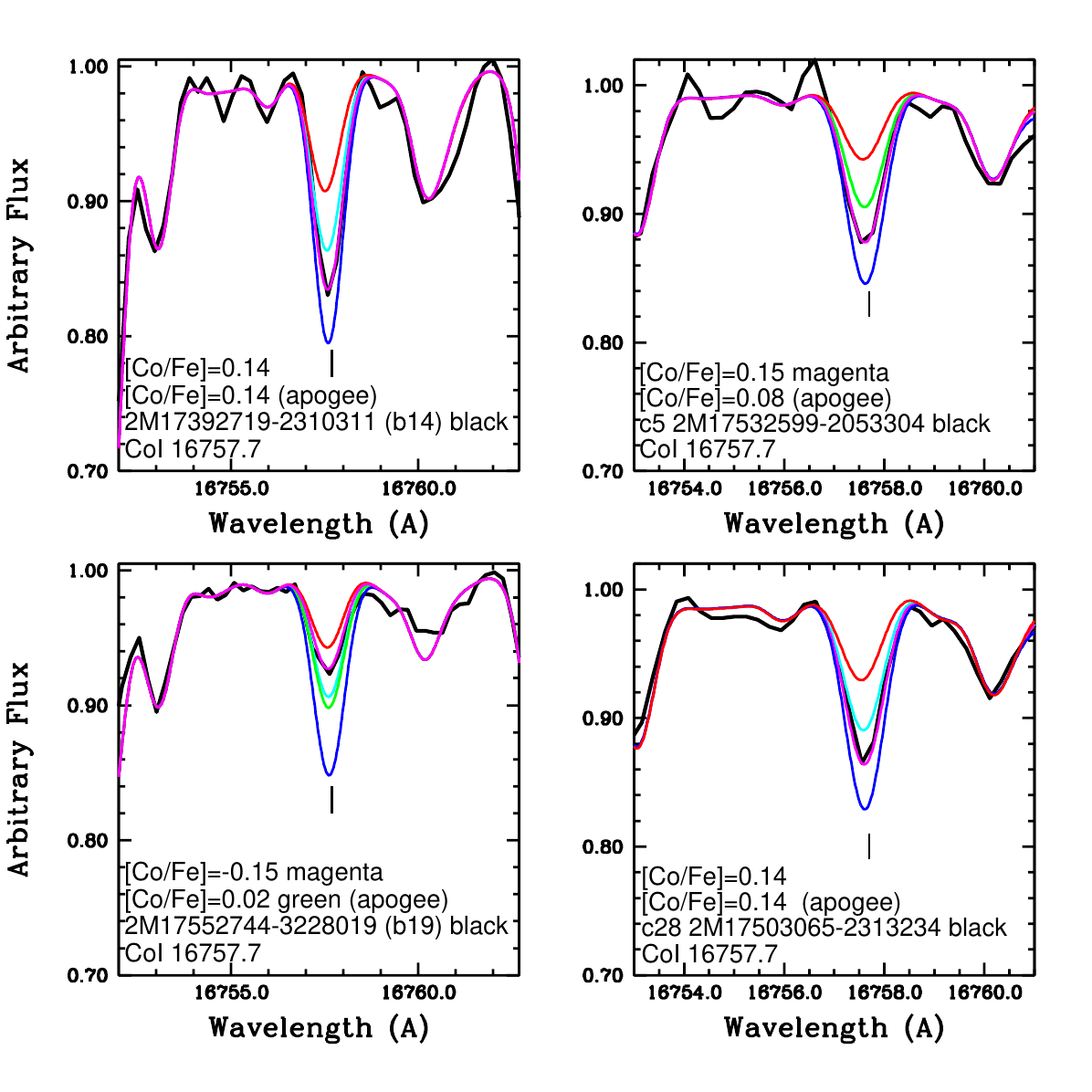}
    \caption{Cobalt line in four stars: 
    2M17392719-2310311 (b14), 2M17552744-3228019 (b19), 2M17532599-2053304 (c5), and 2M17503065-2313234 (c28).
    The synthetic spectra were computed with
    [Co/Fe]=$-$0.3 (red), 0.0 (cyan), +0.3 (blue), the APOGEE
    value indicated in the panels (green), and
    the final value (magenta), and are compared with the observed
    spectrum (black).}
    \label{figco}
\end{figure}

{\it Cobalt:} \ion{Co}{I} 16757.7 {\rm \AA}  is the unique suitable line,  
as listed in \citet{smith13}, but is a very strong and clean line.
We also inspected two other lines, 
CoI 15906.075 {\rm \AA} and 16568.649 {\rm \AA,} but these were not useful. 
Figure \ref{figco} shows the fit to the Co line in four stars, namely
2M17392719-2310311 (b14), 2M17552744-3228019 (b19), 2M17532599-2053304 (c5), and 2M17503065-2313234 (c28).

\begin{figure}
    \centering
    \includegraphics[width=8.5cm]{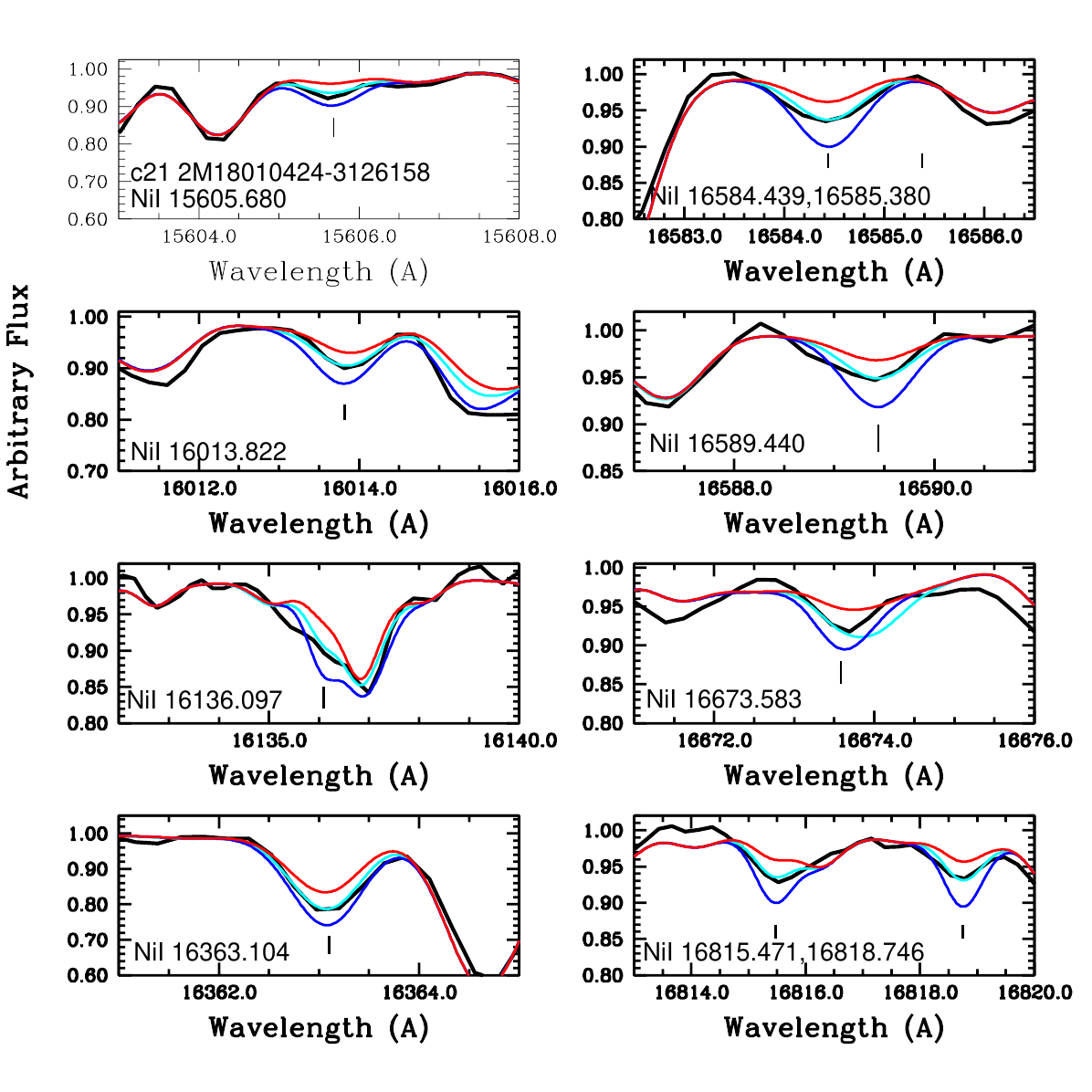}
    \caption{Nickel lines in star 2M18010424-3126158 (c21).
    The synthetic spectra were computed with
    [Ni/Fe]=$-$0.3 (red), 0.0 (cyan), and +0.3 (blue), and are
    compared with the observed  spectrum (black).}
    \label{figni}
\end{figure}

{\it Nickel:}  We analysed nine lines, identified with an asterisk in Table \ref{linelist}, and in general they give the same resulting
abundance. For the cases where the result differs from line to line,
we adopted a mean Ni abundance. The line \ion{Ni}{I} 16589.440 {\AA}
tends to give higher abundances in most stars.
Figure \ref{figni} shows the fit to the nine lines for star
2M18010424-3126158 (c21).

\begin{figure}
    \centering
    \includegraphics[width=8.5cm]{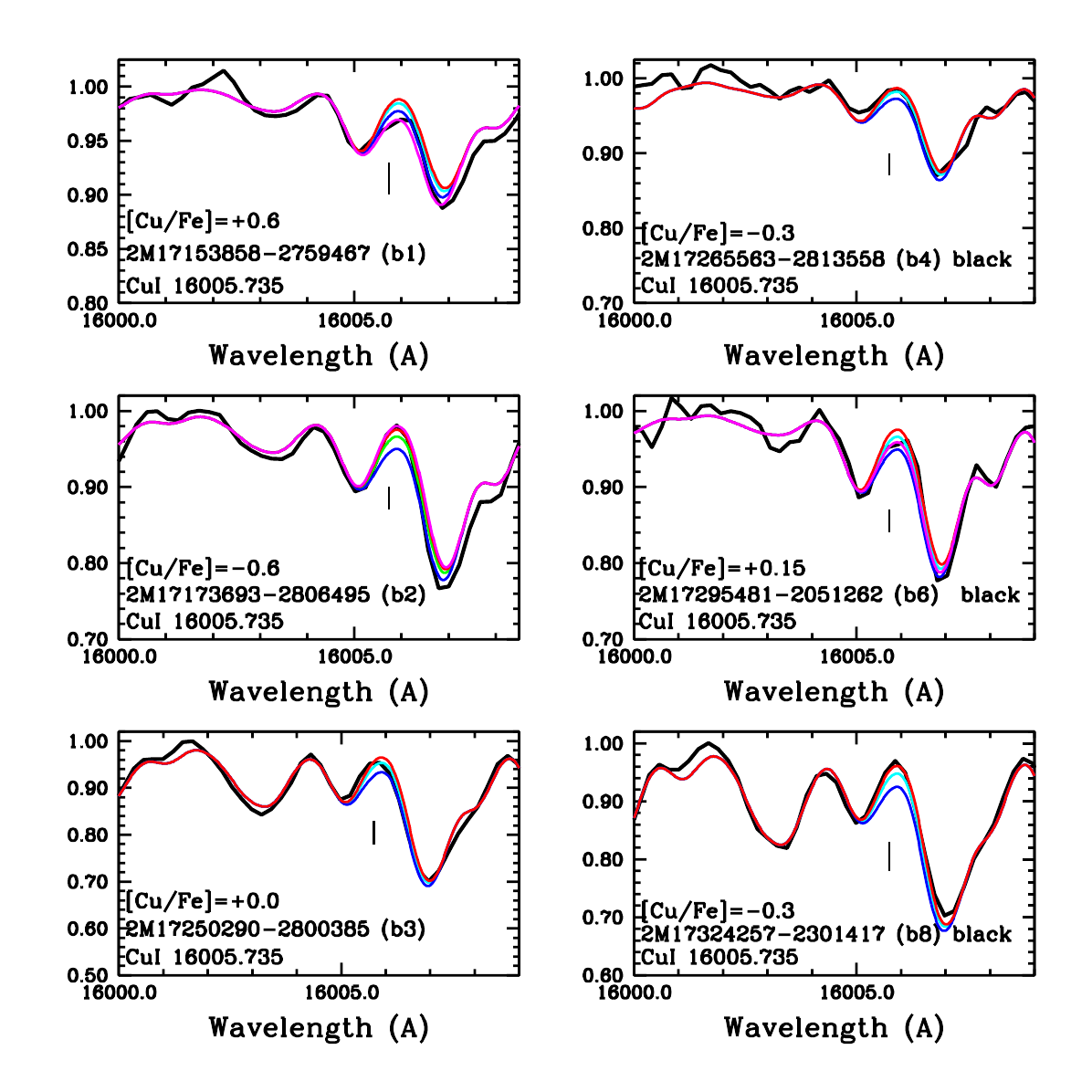}
    \caption{Copper line in six stars illustrating
    the difficulty in deriving Cu abundances.
    The synthetic spectra were computed with
    [Cu/Fe]=$-$0.3 (red), 0.0 (cyan), +0.3 (blue), and
    +0.08 (magenta), an are compared with the observed
    spectrum (black).}
    \label{figcu}
\end{figure}

{\it Copper:}
The CuI 16005.735 {\rm \AA} line identified by \citet{smith13} is the single
useful line.
We also identified the CuI 16650.0 {\rm \AA} line, but 
the latter is immersed  in a strong feature composed of other blending lines; this feature is insensitive to the Cu abundance.
The \ion{Cu}{I} 16005.735 {\rm \AA} line is located between
a line consisting mainly of \ion{Fe}{I}, and a small 
contribution of \ion{Ti}{I}, \ion{Ti}{II}, and \ion{Ca}{I} 
lines, and on the red side a strong \ion{Fe}{I} line.
The \ion{Cu}{I} line is on the wing of the bluer
feature, and can be checked. The main uncertainty in the Cu abundance is due to the adopted
continuum level.
Regarding the continuum, we verified the overall fit to the continuum in the region 16002-16011 {\rm \AA}
and for most stars we used 
the local continuum at around 16009 {\rm \AA\ situated} after the blend dominated by
the \ion{Fe}{I} 16006.758 {\rm \AA} line.
It is also difficult to measure the extent of the Cu deficiency in the severely Cu-deficient cases. 
Figure \ref{figcu} shows the fit to the \ion{Cu}{I} 16005.735 {\rm \AA} line for six stars:
2M17153858-2759467 (b1), 2M17173693-2806495 (b2), 2M17250290-2800385 (b3), 
2M17265563-2813558 (b4), 2M17295481-2051262 (b6), and 2M17324257-2301417 (b8).

\subsection{Non-LTE corrections}

Non-local thermodynamic equilibrium (NLTE) corrections are available for some of the 
lines of Cr, Mn, and Co by \citet{bergemann10}, \citet{bergemann08Mn}, and
\citet{bergemann10Co}, provided on their
website\footnote{https://nlte.mpia.de/gui-siuAC\_secE.php}.
These corrections are reported in Table \ref{nlte}.

We conducted a detailed examination of the NLTE corrections provided on the MPIA website for the available spectral lines. For \ion{Cr}{I}, corrections are available for the lines at 15680.063, 15860.214, and 
16015.327  {\rm \AA}, with an average correction of $\sim$0.15 dex. An inspection of the curve of growth and the line profiles did not reveal any significant discrepancies. However, there are no corrections for the line 
15178.593 {\rm \AA}, which is one of the only two adopted; in addition, the corrections appear too high,
and so we did not apply these NLTE corrections to Cr.

For Mn~I, corrections are provided for the lines at 15217.793 and 15262.702  {\rm \AA}. The line at 15217.793  {\rm \AA} exhibits an average correction of $\sim$0.10 dex, whereas the 15261.702  {\rm \AA} line appears too weak under NLTE conditions, resulting in a larger correction of $\sim$0.5 dex. Therefore, as
not all three lines have corrections available, and  because the corrections appear reasonable for only one line  and high for the rest,
we also did not apply NLTE corrections for Mn.
The corrections on \ion{Co}{I}  16757.711  {\rm \AA} abundances appear suitable, and lead to the finding that  Co appears under-abundant at the lower metallicities.

Finally, the papers cited above as used for the NLTE corrections adopt the classical inelastic collisions with the hydrogen atoms by \citet{drawin68}  in
their calculations. The
approximation  is scaled by a factor S$_{H}$, which in the Bergemann et al. papers is assumed
to have a very low value (S$_{H}$ = 0 for Cr and Mn and S$_{H}$ = 0.05 for Co), thereby
maximising the size of the NLTE corrections. 
For this reason, we finally did not apply the NLTE corrections to the resulting spectroscopic abundances to any of the analysed elements.

\subsection{Uncertainties}

\begin{figure}
    \centering
    \includegraphics[width=8.5cm]{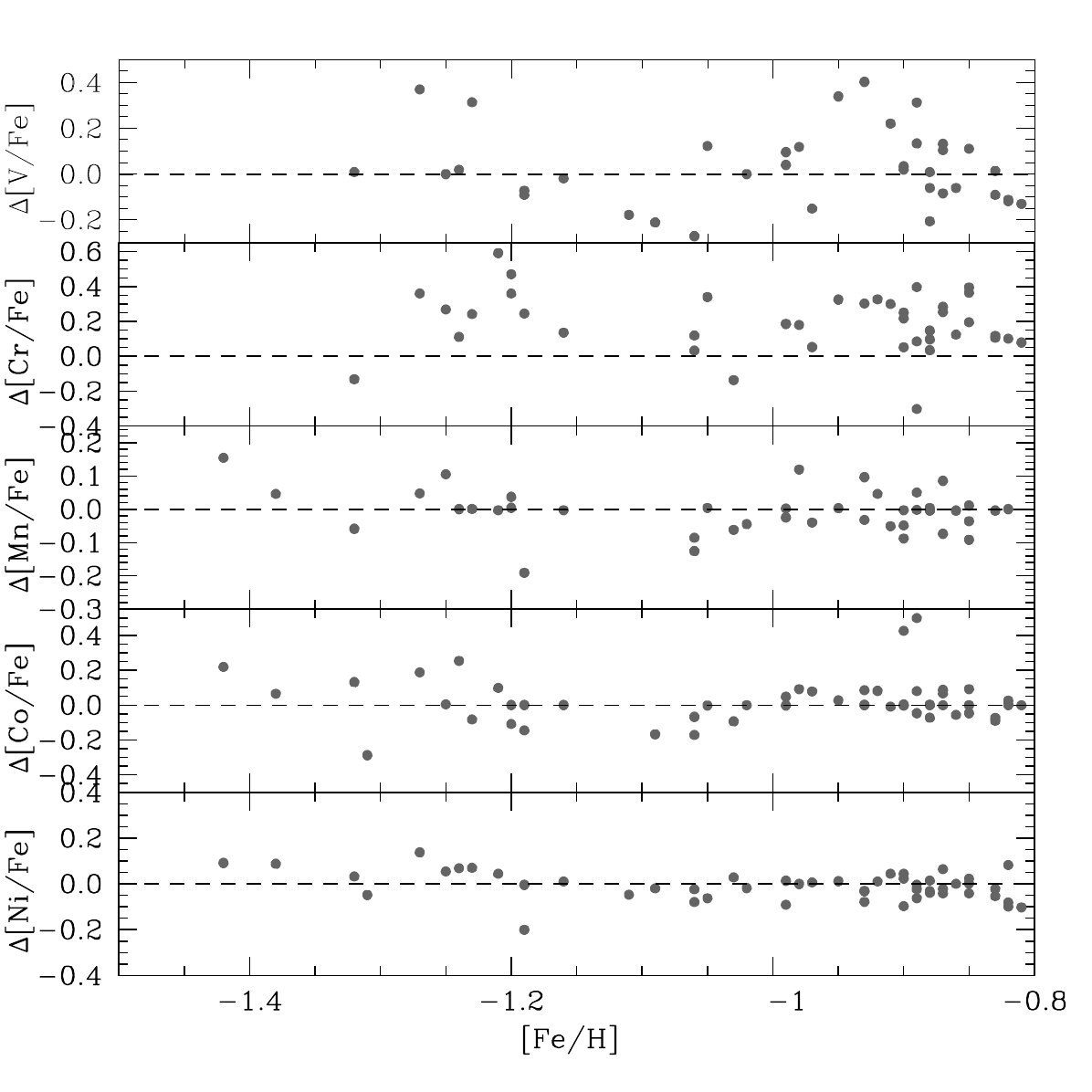}
    \caption{[V, Cr, Mn, Co, Ni/Fe] vs. [Fe/H], plotting the differences between the present results and  the APOGEE DR17 values.}
    \label{plotdiff}
\end{figure}

The uncertainties in the derived abundances can be seen in Figure \ref{plotdiff}, where we plot the difference in abundances, that is, the abundances 
derived in this work minus the ones derived with the APOGEE-ASPCAP DR17 software.
The differences between the two measurements can be considered as the uncertainty, which are mainly due to continuum placement.

There is good agreement between the two values for V, Mn, and Co, with V tending to be higher in the present work, and very good agreement regarding the Ni abundances. Finally, the present
Cr abundances are systematically higher than the ASPCAP ones.

The mean difference between our results and ASPCAP are found to be:

\begin{equation} \label{eq1}
\begin{split}
     {\rm [V/Fe]}_{\rm present}-{\rm [V/Fe]}_{\rm ASPCAP}=+0.11\pm0.005, \\
     {\rm [Cr/Fe]}_{\rm present}-{\rm [Cr/Fe]}_{\rm ASPCAP}=+0.17\pm0.003,\\
     {\rm [Mn/Fe]}_{\rm present}-{\rm [Mn/Fe]}_{\rm ASPCAP}=-0.04\pm0.001,\\
     {\rm [Co/Fe]}_{\rm present}-{\rm [Co/Fe]}_{\rm ASPCAP}=+0.03\pm0.001,\\
     {\rm [Ni/Fe]}_{\rm present}-{\rm [Ni/Fe]}_{\rm ASPCAP}=+0.00\pm0.0002.
\end{split}
\end{equation}

The visual fits are more reliable than ASPCAP for lines in noisy spectra or where lines are
too faint, in which case ASPCAP tends to assign very low abundances, and this
is the reason for our abundances being higher by +0.11 and +0.17 dex for V and Cr.
The VAC-BAWLAS results for vanadium partly mitigate this issue.
For Mn, Co, and Ni, the mean difference is low, but the present results appear more
homogeneous.
The fits are available upon request.

\section{Chemical-evolution models}\label{sec5}

The chemical-evolution model for the Galactic bulge  computed here assumes a classical
bulge. It is derived from the chemical-evolution models of \citet{friaça98},
which adopted a multi-zone chemical evolution
coupled with a hydrodynamical code. For the Galactic bulge, a
classical spheroid with a baryonic mass of 2$\times$10$^{9}$ M$_{\odot}$
and a dark halo mass of 1.3$\times$10$^{10}$ M$_{\odot}$ are assumed
\citep[e.g.][]{barbuy18a}.

Cosmological parameters from the \citet{planck20} are adopted, namely
$\Omega_{m}$ = 0.31, $\Omega_{\Lambda}$ = 0.69, 
Hubble constant H$_{0}$= 68 km s$^{-1}$Mpc$^{-1}$,
and an age of the Universe of 13.801$\pm$0.024 Gyr.

For the nucleosynthesis yields, we adopt:
(i) for massive stars,
the metallicity-dependent yields from core-collapse supernovae (CCSNae)/Supernovae Type II (SNe II) from \citet{woosley95}, with some alterations of the yields
following the suggestions of \citet{timmes95},
and for low metallicities (Z $<$ 0.01 Z$_{\odot}$ , or [Fe/H] $< -$2.5, the yields
are from high-explosion-energy hypernovae (HNe) from \citet{nomoto13};
(ii) Type Ia Supernovae (SNIa) yields are from \citet{iwamoto99} – their models W7 
(progenitor star of initial metallicity
Z = Z$_{\odot}$) and W70 (zero initial metallicity); and (iii) for intermediate-mass stars (0.8–8 M$_{\odot}$) with initial Z = 0.001, 0.004, 0.008, 0.02, and
0.4, we adopt yields from \citet{vandenhoek97}  with variable $\eta$ asymptotic giant branch (AGB case).
Models are computed for radii of r $< 0.5$, $0.5 <$ r $< 1$, $1 <$ r $< 2$,
and $2 <$ r $< 3$ kpc from the Galactic centre, and for specific star-formation rate
values of $\nu = 1$ and $3$ Gyr$^{-1}$.
We note that for Co, the SNe II yields from \citet{woosley95} were computed both with original values and also multiplied by two, following the recommendations of Timmes et al. (1995), see further discussion below. The yields of SNe Ia for Ni from \citet{iwamoto99} were divided by two, because, as the authors pointed out, the W7 model overproduces $^{58}$Ni, the main Ni isotope. More over, from Table 3 of \citet{iwamoto99}, the W7 model has [$^{58}$Ni/$^{56}$Fe]=+0.62, and the W70 model, [$^{58}$Ni/$^{56}$Fe]=+0.47, whereas the five remaining models have an average [$^{58}$Ni/$^{56}$Fe]=+0.14,  implying $^{58}$Ni overabundances by factors of $\sim$ 3 and  $\sim$ 2, for models W7 and W70, respectively.



\begin{figure}
    \centering
    \includegraphics[width=9.0cm]{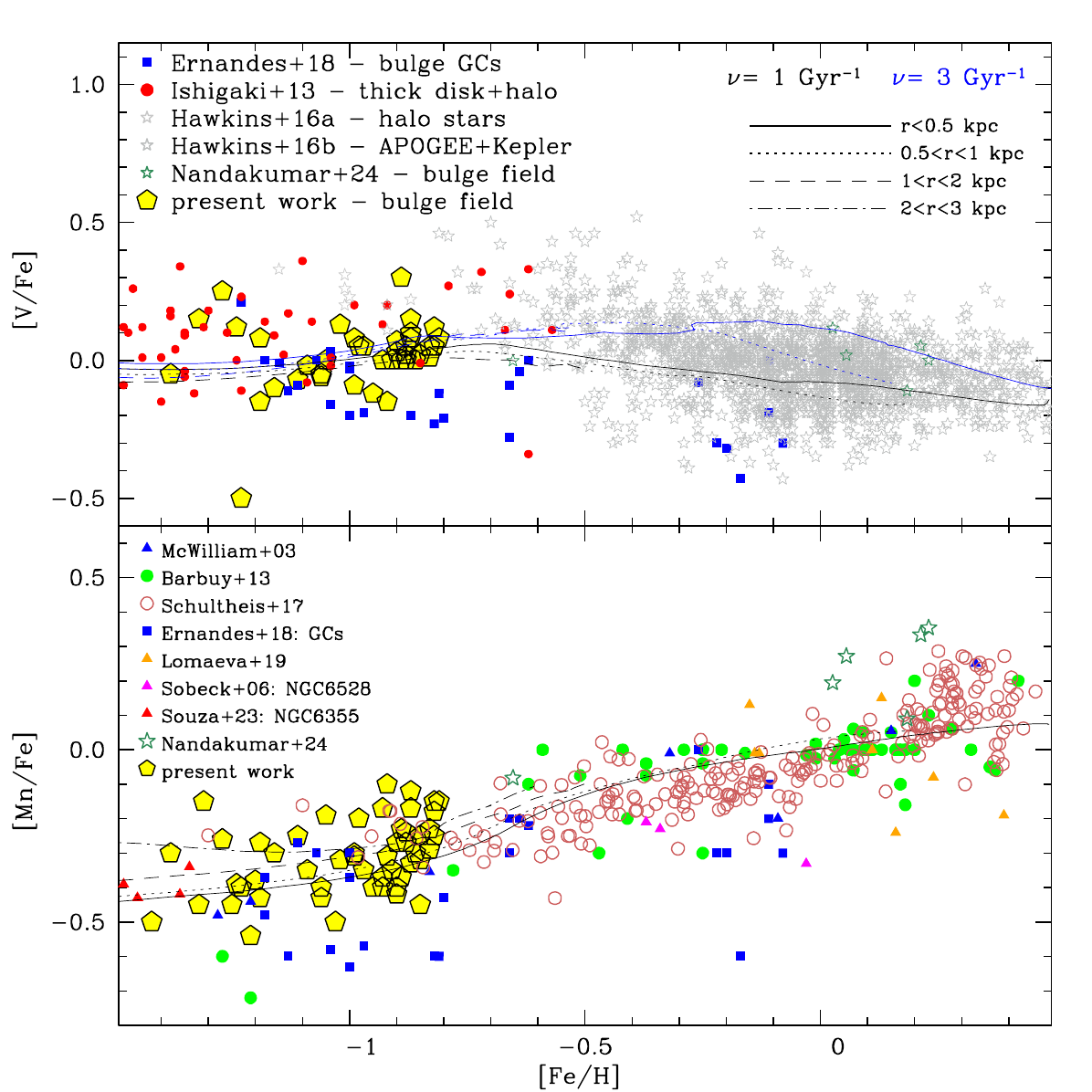}
    \caption{[V/Fe] vs. [Fe/H] (upper panel) and [Mn/Fe] vs. [Fe/H] (lower panel): Chemical-evolution models with SFR of $\nu$ = 1  and 3 Gyr $^{-1}$ (black and blue lines respectively) overplotted on the results of the present  study (yellow pentagons)  and  literature data.
    Bulge GCs are from \citet{ernandes18} (blue squares), 
    bulge field stars are from \citet{nandakumar24} (sea green open stars). For V:
 thick disc and halo stars from \citet{ishigaki13} (red filled circles),  
    halo stars from \citet{hawkins16a}, and
    disc stars from Kepler+APOGEE given in \citet{hawkins16b} (bold grey open stars).
    For Mn: bulge field data from  \citet{lomaeva19} (filled orange triangles),  \citet{schultheis17} (open red circles), \citet{barbuy13} (filled green circles), \citet{mcwilliam03} (filled blue triangles), and GCs NGC~6528 by \citet{sobeck06} (filled magenta triangles), NGC~6355 by \citet{souza23} (filled red triangles).  
    Different model lines correspond to the outputs of models computed for radii r $<$ 0.5, 0.5 $<$ r $<$ 1, 1 $<$ r $<$ 2, and 2 $<$ r $<$ 3 kpc from the Galactic centre.} 
    \label{plotv}
\end{figure}

\begin{figure}
    \centering
    \includegraphics[width=9.0cm]{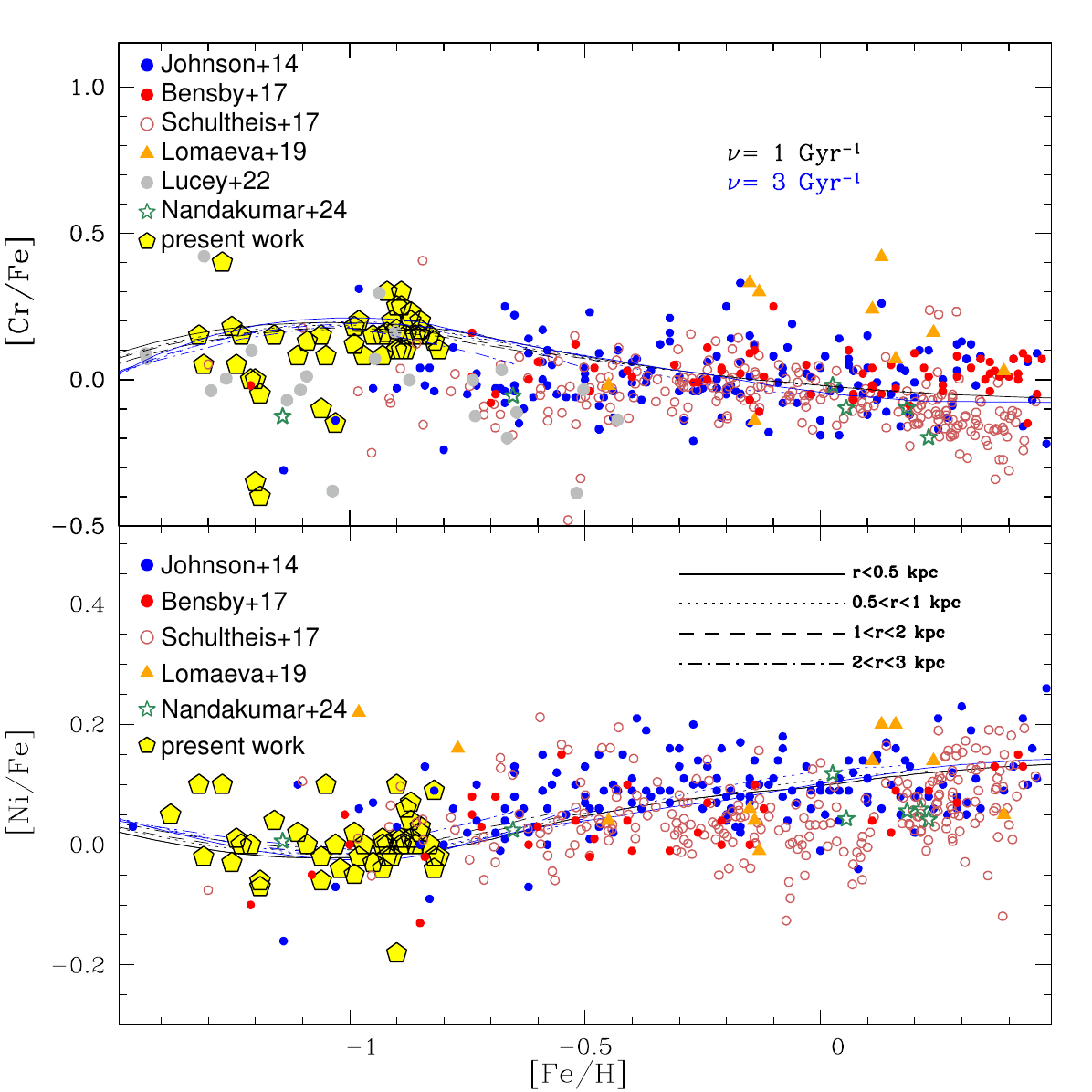}
    \caption{ [Cr/Fe]  vs. [Fe/H] (upper panel)  and [Ni/Fe] vs. [Fe/H]  (lower panel).   Chemical-evolution models with SFR of $\nu$ = 1  and 3 Gyr $^{-1}$ (black and blue lines respectively) overplotted on the results of the present study (yellow pentagons)  and  literature data: 
    \citet{johnson14} (blue filled circles), \citet{bensby17} (red filled circles), \citet{schultheis17} (open red circles), \citet{lomaeva19} (filled orange triangles), and \citet{lucey22} (grey filled circles, only for Ni).
    Different model lines are for the same radii from the Galactic centre as in Fig. \ref{plotv}.
     } 
    \label{plotcrni}
\end{figure}

\begin{figure}
    \centering
    \includegraphics[width=9.0cm]{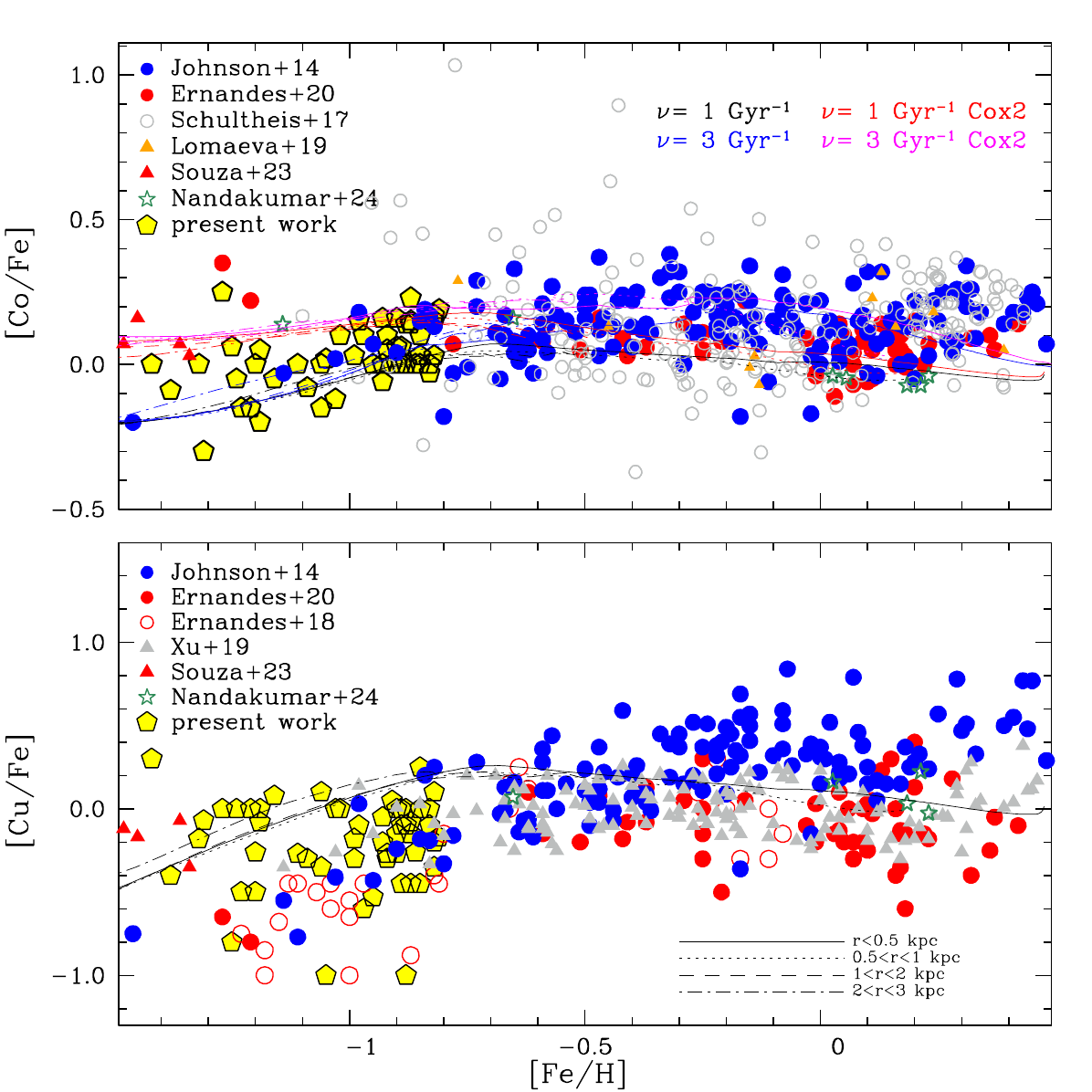}
    \caption{[Co/Fe]  vs. [Fe/H] (upper panel) and [Cu/Fe] vs. [Fe/H] (lower panel):  Chemical-evolution models with specific star formation of $\nu$ = 1 and 3  Gyr $^{-1}$ 
      for original yields from \citet{woosley95} (black and blue respectively), and for Co yields multiplied by two (red and magenta lines). Data consist of the present results (yellow pentagons) and  literature data, including  field data from \citet{johnson14} (blue filled circles), \citet{ernandes20} (red filled circles),   and \citet{schultheis17} (open grey circles), and  \citet{xu19} (filled grey triangles), plus GC data from  \citet{ernandes18} (red open circles) in the lower panel. Different model lines are for the same radii from the Galactic centre, as in Fig. \ref{plotv}.} 
    \label{plotcocu}
\end{figure}

Below we describe the available literature data for the studied elements, which are taken into account in
the [X/Fe] versus [Fe/H] plots, and the chemical-evolution models.
Figures  \ref{plotv}, \ref{plotcrni}, and 
\ref{plotcocu} 
show [V/Fe] versus [Fe/H] (upper panel) and 
[Mn/Fe] versus [Fe/H] (lower panel), [Cr/Fe] versus [Fe/H] (upper panel) and [Ni/Fe] versus [Fe/H] (lower panel), and [Co/Fe] versus [Fe/H] (upper panel) and [Cu/Fe] versus [Fe/H] (lower panel), respectively.
The present data are compared with literature data
and the chemical-evolution models described above, that is, for specific star formation rates
of $\nu_{SF}$ = 1 and 3 Gyr$^{-1}$.
The literature data included in these plots are described below.

{\it Vanadium}: In Figure \ref{plotv} (upper panel) the present data are
compared with the following literature data: 
bulge GCs from \citet{ernandes18},
    thick-disc and halo stars from \citet{ishigaki13}, 
     halo stars from \citet{hawkins16a}, disc stars from Kepler+APOGEE given in \citet{hawkins16b},     and bulge field stars by
    \citet{nandakumar24}. As most of the available data on V do not correspond to
    bulge stars, the origin of the different samples is indicated in the figure panel.
    The data show a large spread. 

{\it Manganese}: In Figure \ref{plotv} (lower panel)  the present data are compared with literature data, including the  GCs from \citealt{ernandes18}, namely the metal-rich bulge clusters
NGC~6528 and NGC~6553, the moderately metal-poor clusters HP~1, NGC~6522, and NGC~6558, and additionally the disc cluster 47 Tucanae; again results for the metal-rich cluster NGC~6528 from \citet{sobeck06}, and the  relatively metal-poor cluster NGC~6355 from \citet{souza23}, and bulge field stars 
 \citep{nandakumar24,lomaeva19,schultheis17,barbuy13,mcwilliam03}.
We note that \citet{schultheis17} are early results from APOGEE Data Release 13 (DR13).
 Mn abundances from  \citet{lucey22} are not plotted because 
of the large scatter for this element. 
The models appear to fit the data very well, however we note that the NLTE corrections were
not taken into account.

Manganese behaves as a metallicity-dependent element. [Mn/Fe] decreases with decreasing metallicity due to the decreasing trend in the yields  from  SNe II ejecta at metallicities of lower than [Fe/H] $\sim$ $-$1. In addition, for solar metallicities, the SNe II Mn yields have almost solar abundances and the SNIa contribution becomes noticeable for [Fe/H]$>-$0.8. For the W70 model of \citet{iwamoto99}, [Mn/Fe] = $-$0 07, and [Mn/Fe] = +0.10 for the W7 model. The impact of the SNe Ia is clearly seen in Figure \ref{plotv}, in the small jump of [Mn/Fe] at [Fe/H] $\sim$ $-$0.8, as shown by the chemodynamical model with $\nu_{SF}$ = 3 Gyr$^{-1}$.

{\it Chromium:} Literature data include bulge field stellar abundances \citep{nandakumar24,lucey22,lomaeva19,bensby17,schultheis17,johnson14}.
Cr tends to vary in lockstep with Fe.
The models for Cr show a suprasolar bump at [Fe/H]$\sim$-1.1 and a subsolar bow at [Fe/H]$\sim$-2.0 as a
direct consequence of the yields from \citet{woosley95}.
Again, the data presented by \citet{schultheis17}  are early results from APOGEE DR13.

{\it Cobalt:} Literature data include bulge field-star \citep{nandakumar24,ernandes20,lomaeva19,schultheis17,johnson14}. Co tends to vary in lockstep with Fe, but could be somewhat under-abundant in the more metal-poor stars, which is not reproduced by the models.
 In Figure \ref{plotcocu}, the NLTE Co abundances are 
taken into account in results from \citet{ernandes20}, but this is  not the case for the present data due to the reasons explained in subsection 4.1.
For Co, the data from \citet{schultheis17} are early results from APOGEE DR13.
The adopted yields from \citet{woosley95} are shown with original values, but also
with Co yields multiplied by two, as recommended by \citet{timmes95}. It appears that with
the new data for the more metal-poor stars, the original yields are more suitable.

{\it  Nickel:} Literature data include bulge field-star \citep{nandakumar24,lomaeva19,bensby17,schultheis17,johnson14}. Regarding the data from \citet{bensby17}, we adopt only the stars older than 11 Gyr. 
\citet{schultheis17} are early results from APOGEE DR13.
Ni clearly varies in lockstep with Fe. Relative to the other elements studied here, the models for Ni follow the Fe  more closely, with a deviation from solar ratios 
at around [Ni/Fe]$< \pm$ 0.1. This applies to the models and also to the data.

{\it  Copper:} Literature data include  bulge field-star abundances \citep{nandakumar24,ernandes20,lomaeva19,xu19,johnson14}, and GC data are from \citet{ernandes18}.

In conclusion, Cr and Ni vary in lockstep with Fe, that is, [Cr/Fe] $\sim$ [Ni/Fe] $\sim$ 0.0 for all metallicities. Mn is deficient for metal-poor stars and shows a clear secondary behaviour.
 Regarding Co and Cu, in \citet{barbuy18a} and \citet{ernandes20}, we suggested that abundances 
 of Co and Cu  could be used to discriminate between their origin as 
neutron capture elements on iron-group nuclei during He burning and later
burning stages, also called the weak-$s$ component \citep{limongi03}, 
and the $\alpha$-rich freeze-out in the deepest layers
\citep{woosley73,woosley95, woosley02, sukhbold16}. 
The results of \citet{ernandes20} seemed to clearly show that Co follows Fe, but we only a few points at [Fe/H]$\leq -1$ were available to these authors. With the present data, we see that for some stars, Co tends to vary in lockstep with Fe, showing [Co/Fe]$\sim 0.0$, but there are also several stars with a lower [Co/Fe] for [Fe/H]$<-$1. 
As mentioned above, following a suggestion by \citet{timmes95}, the yields of Co from
\citet{woosley95} multiplied by two satisfy the higher Co values, but for the Co-low more metal-poor stars, the original lower Co yields provide a better fit to the data.
This is now evidence that Co, and likewise Cu, might have a secondary behaviour, not being dominantly produced by alpha-rich freeze-out as we suggested in \citet{ernandes20} but by a weak-$s$ process.
Cu shows a clear secondary behaviour, but
another issue is that Cu appears to be even more deficient than predicted by the models for [Fe/H]$<-$0.8.

Finally, \citet{zasowski19} studied the APOGEE DR14 and DR15 abundances of 4000 stars
located at distances of $<$ 4 kpc from the Galactic centre, including Cr, Mn, Co, and Ni.
The data naturally show a large spread, but the general trend of Cr, Mn, and Ni is compatible
with the present results, whereas their Co abundances show too large a spread to be compared with the present results.



\section{Inspecting abundance indicators}\label{sec6}

Table \ref{all} lists the abundances of elements studied in \citet{razera22} and
\citet{barbuy23}. Together with 
Table \ref{results}, these data gather all abundances verified by our calculations,
in addition to that of cerium, which was also analysed in \citet{razera22}.

We have analysed some of these abundances as indicators of origin as second generation
from GCs, and abundance indicators of in situ or ex situ origin.

\begin{figure}
    \centering
    \includegraphics[width=8.5cm]{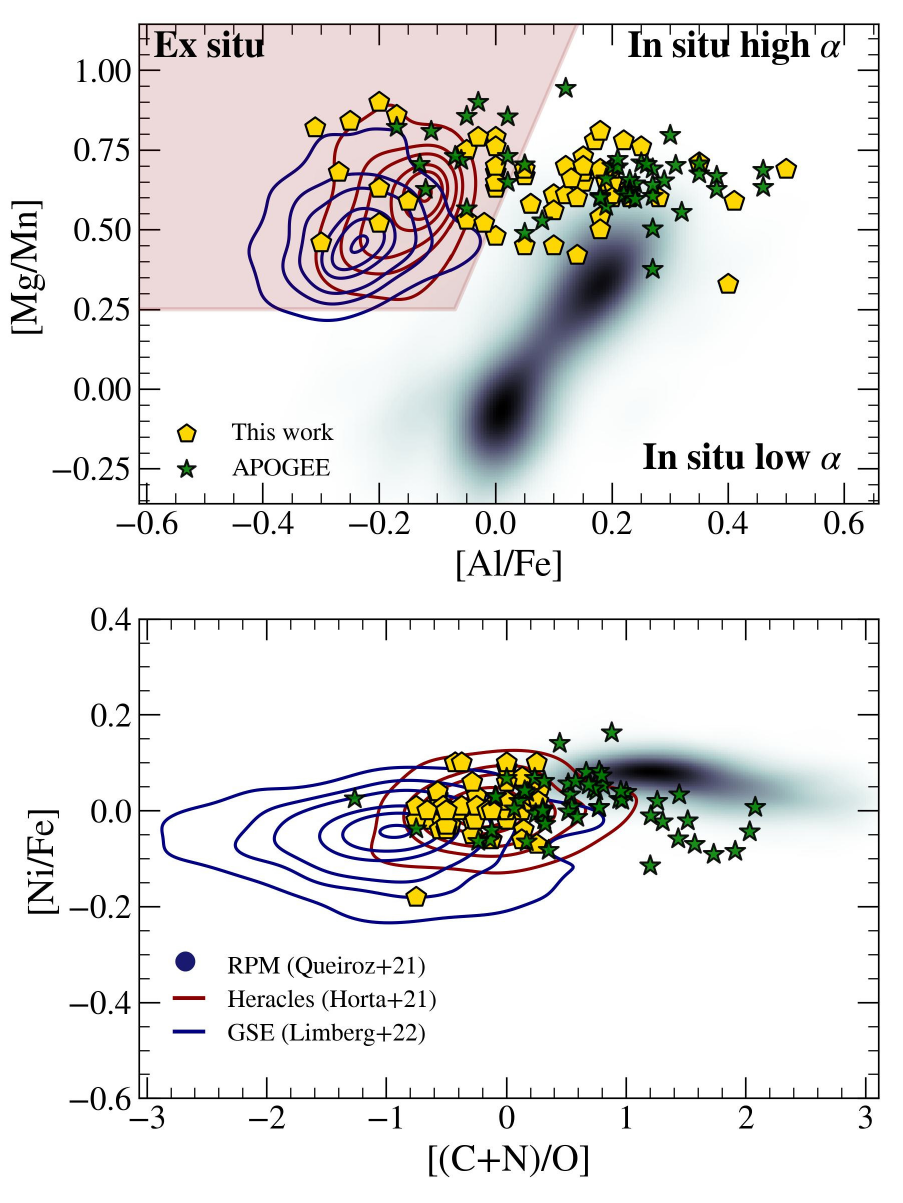}
    \caption{[Mg/Mn]  vs. [Al/Fe] (upper panel) and [Ni/Fe] vs. [C+N/O] (lower panel). Contours correspond
    to Heracles \citet{horta21} and Gaia-Sausage-Enceladus \citet{limberg22}, and the black points to the
    RPM sample from \citet{queiroz21}. Yellow pentagons show abundances derived from the results presented here, while blue stars show  APOGEE-ASPCAP abundances for the same 58 stars.
    } \label{insitu}
\end{figure}

\begin{itemize}

\item Nitrogen-rich stars in the bulge field were identified as second-generation stars evaporated from GCs by \citet{schiavon17} and \citet{fernandez-trincado17}.
In our sample stars,  b15, c11, and C13 
are N-rich with [N/Fe]$>$+0.5, and could be considered as second-generation stars evaporated from GCs.
The star c13 also has a rather high Na abundance of [Na/Fe]=+0.35; see Table \ref{all}.

\item We apply the correlations discussed in
\citet{montalban21},and \citet{ortigoza23}, who proposed that [Ni/Fe] versus [(C+N)/O] is an indicator that can be used to discriminate between an in situ and an ex situ stellar origin, similar to the more popular representation in the [Mg/Mn] versus [Al/Fe] plane. We show both the present abundances and
the APOGEE-ASPCAP abundances for the same 58 stars.
Figure \ref{insitu} shows these plots, where we can see that in the [Mg/Mn] versus [Al/Fe] plane there are stars spread between the in situ and the ex situ region.
The difference with the \citet{queiroz21}  population is larger for the results
in this study when compared with the DR17 results that looked more spread between
the two regions.
In the [Ni/Fe] versus [(C+N)/O] plot, more stars fall in the Heracles locus. For [Mg/Mn] versus [Al/Fe], \citet{queiroz21} demonstrated that 
counter-rotating stars populate both the in situ and ex situ regions. \citet{vasini24} demonstrate that the uncertainties on the chemical yields can reproduce totally different star formation histories (SFHs) in the [Mg/Mn] versus [Al/Fe] plane, which means that they may reproduce an ex situ SFH using in situ yields. On the other hand, \citet{Feltzing23} argue
that this kind of plot can lead to a clear separation between in situ metal-weak disc 
and halo ex situ stars, which otherwise also show a high and moderate [Mg/Fe], respectively.
As for Ni, subsolar Ni is expected for ex situ stars, whereas [Ni/Fe] = 0.00±0.05 for our sample, which is characteristic of an in situ origin. Second-generation stars can also mimic an in situ origin on the [Mg/Mn] versus [Al/Fe] plane, as observed by \citet{fernandez-trincado22a} and
\citet{souza23}, who analysed individual stars of in situ GCs. For instance, the GC NGC~6558, classified as in situ based on different methods \citep{barbuy18b,perez-villegas20,souza24}, has stars with low [Al/Fe], which could lead to a misinterpretation that NGC~6558 is an ex situ cluster, if only this criterion were taken into account.

\item \citet{nissen24} point out a parallel between the abundances of alpha elements and iron-peak elements.
For their alpha-low stars, V, Sc, and Co are also low. Our stars are all alpha-enhanced, as can be seen from their O and Mg abundances, and somewhat lower values of Ca and Si - this is expected given that O and Mg are produced in hydrostatic phases of massive stars, whereas Ca and Si are produced in explosive phases \citep{mcwilliam16}.
For Nissen's stars, the alpha-enhanced stars show [Co/Fe]$\sim$+0.1,
like most of our stars, and the alpha-low ones show [Co/Fe]$\sim -$0.1. 
Also, \citet{nissen10} found that
nearly all accreted stars in the solar neighbourhood have  $-$0.2 $<$ [Ni/Fe] $<$ $-$0.05.
Our sample stars are all alpha-enhanced and most of them also show [Co/Fe]$\sim$+0.0
to 0.1, and [Ni/Fe] $\sim$ 0.0,
therefore indicating an in situ origin.

\end{itemize}

In conclusion, we believe that [(C+N)/O] does not seem to offer a robust way of discriminating between an ex situ and an in situ origin, as it 
contradicts the other more reliable indicators such as alpha-enhancement, but also [Al/Fe], which is low
in Heracles stars ($-$0.25 $\simless$ [Al/Fe] $\simless$ 0.0).   CNO also depends
critically on stellar evolution processes of mixing and extra-mixing 
\citep[e.g.][]{smiljanic09,shetrone19}, whereas
the other abundance indicators depend on the nucleosynthesis that took place in the environment where
the stars were born.

As Heracles selection criteria include high eccentricities and lower energies, some overlap with our selection criteria for selecting our RPM sample
\citep[see][]{queiroz20,queiroz21} is expected..

There is coincidences with Heracles, with however  one discriminator being the low [Al/Fe] in Heracles stars. 
This structure might well be an early progenitor of the early bulge of the Milky Way, and not
an accreted object.
A similar case is Aurora, which shows a wide range in [Al/Fe], and was considered by \citet{myeong22} 
to contain early stars formed in the young Milky Way.

\begin{table*}
\centering
\caption[5]{Abundances of C, N, and O \citep{razera22}, Na and Al \citep{barbuy23}, and Mg, Si, and Ca from DR17.}
\resizebox{0.9\textwidth}{!}{
\begin{tabular}{lcc | ccc | cc | c | ccc}
\noalign{\smallskip}
\hline
\noalign{\smallskip}
\hbox{ID} & \hbox{ID}  & [Fe/H] & [C/Fe] & [N/Fe] & [O/Fe] & [Na/Fe] & [Al/Fe] & [Al/Fe] & [Mg/Fe] &  [Si/Fe] & [Ca/Fe] \\  
int. & 2MASS & DR17 & \multicolumn{3}{c}{Razera+22} &  \multicolumn{2}{c}{DR17} &  Barbuy+23 & \multicolumn{3}{c}{DR17}   \\
\hline
\noalign{\smallskip}
b1  &  2M17153858-2759467       &  $-$1.65 & $-$0.60 & +0.40 & +0.35 & +0.25          & $-$0.13 & $-$0.27  &  0.33  &   0.13  &   0.19       \\
b2  &  2M17173693-2806495       &  $-$0.97 & $-$0.20 & +0.00 & +0.40 & $-$0.25        & +0.19   & +0.12    &  0.35  &   0.22  &   0.22     \\
b3  &  2M17250290-2800385       &  $-$0.82 & $-$0.05 & +0.10 & +0.35 & $-$0.06$^{*}$   & +0.04   & +0.00   &   0.33  &   0.10  &   0.11     \\
b4  &  2M17265563-2813558       &  $-$1.32 & $-$0.35 & +0.20 & +0.35 & +0.69          & +0.02   & +0.00    &  0.34  &   0.31  &   0.21     \\
b5  &  2M17281191-2831393       &  $-$1.19 & $-$0.30 & +0.40 & +0.40 & $-$0.39        & +0.05   & +0.00    &  0.25  &   0.21  &   0.19     \\
b6  &  2M17295481-2051262       &  $-$0.85 & $-$0.30 & +0.20 & +0.40 & $-$0.72        & +0.38   & +0.25    &  0.31  &   0.30  &   0.24     \\
b7  &  2M17303581-2354453       &  $-$0.98 & $-$0.25 & +0.00 & +0.40 & $-$0.20        & +0.20   & +0.06    &  0.38  &   0.19  &   0.20     \\
b8  &  2M17324257-2301417       &  $-$0.82 & +0.00   &$-$0.10 & +0.35 & {\bf +0.25}$^{*}$ & +0.22   & +0.18   &  0.36  &   0.18  &   0.19     \\
b9  &  2M17330695-2302130       &  $-$0.93 & +0.00 & +0.00 & +0.35 & $-$0.16 &$-$0.08  & +0.10   &   +0.28  &   0.14  &   0.26     \\
b10 &  2M17344841-4540171       &  $-$0.88 & $-$0.30 & +0.20 & +0.35 & $-$0.22$^{*}$   & +0.27   & +0.10   &  0.34  &   0.27  &   0.26     \\
b11 &  2M17351981-1948329       &  $-$1.11 & $-$0.10 & +0.10 & +0.40 & $-$0.15        &$-$0.01  & $-$0.05 &   0.28  &   0.06  &   0.21     \\
b12 &  2M17354093-1716200       &  $-$0.87 & $-$0.20 & +0.00 & +0.37 & $-$0.20        & +0.24   & +0.18   &   0.39  &   0.17  &   0.21     \\
b13 &  2M17390801-2331379       &  $-$0.81 & $-$0.10 & +0.15 & +0.38 & {\bf +0.25}$^{*}$ & +0.05   & +0.05   &  0.33  &   0.15  &   0.12     \\
b14 &  2M17392719-2310311       &  $-$0.87 & $-$0.10 & +0.10 & +0.38 & $-$0.15$^{*}$   &$-$0.06  & +0.05   &  0.25  &   0.11  &   0.21     \\
b15 &  2M17473299-2258254       &  $-$1.74 & $-$0.70 & +0.80 & 0.35 & +0.15          &$-$0.11  & $-$0.25 &   0.34  &   0.27  &   0.15     \\
b16 &  2M17482995-2305299       &  $-$1.03 & $-$0.30 & +0.30 & +0.40 & +0.06$^{*}$     &$-$0.06  & $-$0.20 &  0.28  &   0.40  &   0.01     \\
b17 &  2M17483633-2242483       &  $-$1.09 & $-$0.20 & +0.10 & +0.35 & {\bf +0.08}$^{*}$ &+0.08    & +0.06   &  0.28  &   0.13  &   0.25     \\
b18 &  2M17503263-3654102       &  $-$0.99 & $-$0.40 & +0.40 & +0.33 & +0.08$^{*}$     & +0.23   & +0.17   &  0.38  &   0.31  &   0.25     \\
b19 &  2M17552744-3228019       &  $-$1.06 & $-$0.30 & +0.40 & +0.35 & $-$0.08        & +0.23   & +0.10   &   0.31  &   0.32  &   0.20     \\
b20 &  2M18020063-1814495       &  $-$1.38 & $-$0.50 & +0.30 & +0.35 & $-$0.20        &$-$0.05  & $-$0.20:&   0.22  &   0.19  &   0.13     \\
b21 &  2M18050452-3249149       &  $-$1.16 & $-$0.50 & +0.20 & +0.40 & $-$0.45        & +0.23   & +0.14   &   0.32  &   0.35  &   0.24  \\  
b22 &  2M18050663-3005419       &  $-$0.92 & $-$0.10 & +0.00 & +0.40 & +0.01$^{*}$     &$-$0.08  & $-$0.02 &  0.29  &   0.14  &   0.12     \\
b23 &  2M18065321-2524392       &  $-$0.89 & $-$0.20 & +0.20 & +0.38 & $-$0.73        & +0.27   & +0.20   &   0.36  &   0.23  &   0.23     \\
b24 &  2M18104496-2719514       &  $-$0.82 & $-$0.10 & +0.10 & +0.35 & $-$0.05        & +0.25   & +0.12   &   0.46  &   0.20  &   0.16     \\
b25 &  2M18125718-2732215       &  $-$1.31 & $-$0.22 & +0.20 & +0.40 & $-$0.31        & $-$0.21 & $-$0.30 &   0.16  &   0.11  &   0.16     \\
b26 &  2M18200365-3224168       &  $-$0.86 & $-$0.35 & +0.20 & +0.32 & $-$0.05        & +0.35   & +0.22   &   0.38  &   0.27  &   0.21     \\
b27 &  2M18500307-1427291       &  $-$0.95 & $-$0.30 & +0.20 & +0.38 & $-$0.10        & +0.31   & +0.15   &   0.30  &   0.29  &   0.18     \\
c1  &  2M17173248-2518529       &  $-$0.91 & $-$0.25 & +0.20 & +0.38 & $-$0.13$^{*}$   & +0.08   & +0.00   &  0.23  &   0.15  &   0.21     \\
c2  &  2M17285088-2855427       &  $-$1.23 & $-$0.45 & +0.40 & +0.40 & $-$0.38        &$-$0.07  & +0.00   &   0.33  &   0.18  &   0.19     \\
c3  &  2M17301495-2337002       &  $-$1.06 & $-$0.25 & +0.20 & +0.40 & +0.26          & +0.19   & +0.19   &   0.27  &   0.22  &   0.25     \\
c4  &  2M17453659-2309130       &  $-$1.20 & $-$0.30 & +0.30 & +0.40 & $-$0.15        &$-$0.12  & $-$0.15 &   0.21  &   0.15  &   0.11     \\
c5  &  2M17532599-2053304       &  $-$0.87 & $-$0.25 & +0.20 & +0.40 & $-$0.19        & +0.18   & +0.15   &   0.35  &   0.30  &   0.19     \\
c6  &  2M18044663-3132174       &  $-$0.90 & $-$0.15 & +0.00 & 0.40 & $-$0.06        & +0.18   & +0.20   &   0.38  &   0.18  &   0.26     \\
c7  &  2M18080306-3125381       &  $-$0.90 & $-$0.30 & +0.00 & +0.40 & $-$0.70        & +0.46   & +0.50   &   0.29  &   0.35  &   0.26     \\
c8  &  2M18195859-1912513       &  $-$1.24 & $-$0.40 & +0.40 & +0.40 & $-$0.16        & +0.02   & +0.00   &   0.26  &   0.23  &   0.21     \\
c9  &  2M17190320-2857321       &  $-$1.20 & $-$0.30 & +0.20 & +0.70 & +0.53          & +0.05   & +0.00   &  0.32   &  0.28   &  0.21      \\
c10 &  2M17224443-2343053       &  $-$0.88 & $-$0.35 & +0.20 & +0.37 & $-$0.10$^{*}$   & +0.38   & +0.23   & 0.39   &  0.34   &  0.28      \\
c11 &  2M17292082-2126433       &  $-$1.27 & $-$0.60 & +0.60 & +0.38 & +0.21          & +0.27   & +0.40   &  0.07   &  0.16   &  0.28      \\
c12 &  2M17323787-2023013       &  $-$0.85 & $-$0.15 & +0.20 & +0.40 & $-$0.05        & +0.22   & +0.20   &  0.35   &  0.24   &  0.21      \\
c13 &  2M17330730-2407378       &  $-$1.90 & $-$0.70 & +0.70 & +0.38 & +0.35          &$-$0.17  & $-$0.31 &  0.28   &  0.24   &  0.28      \\
c14 &  2M18023156-2834451       &  $-$1.19 & $-$0.05 & +0.10 & +0.40 & $-$0.64        & +0.01   & +0.10   &  0.29   &  0.13   &  0.28      \\
c15 &  2M17291778-2602468       &  $-$0.99 & $-$0.20 & +0.30 & +0.38 & $-$0.44        & +0.27   & +0.20   &   0.38  &   0.30  &   0.15     \\
c16 &  2M17310874-2956542       &  $-$0.93 & $-$0.40 & +0.20 & +0.36 & $-$0.02        & +0.21   & +0.15   &   0.36  &   0.35  &   0.22     \\
c17 &  2M17382504-2424163       &  $-$1.05 & $-$0.20 & +0.30 & +0.40 & +0.26$^{*}$     & +0.27   & +0.18   &  0.31  &   0.17  &   0.13     \\
c18 &  2M17511568-3249403       &  $-$0.90 & $-$0.20 & +0.00 & +0.38 & $-$0.02        & +0.21   & +0.18   &   0.39  &   0.24  &   0.21     \\
c19 &  2M17552681-3334272       &  $-$0.89 & $-$0.30 & +0.00 & +0.40 & $-$0.14        & +0.46   & +0.41   &   0.36  &   0.43  &   0.21     \\
c20 &  2M18005152-2916576       &  $-$1.02 & $-$0.40 & +0.20 & +0.40 & +0.16          & +0.32   & +0.23   &   0.28  &   0.40  &   0.24     \\
c21 &  2M18010424-3126158       &  $-$0.83 & $-$0.25 & +0.00 & +0.38 & {\bf +0.02}$^{*}$ & +0.16   & +0.05   &  0.40  &   0.20  &   0.20     \\
c22 &  2M18042687-2928348       &  $-$1.21 & $-$0.50 & +0.30 & +0.40 & $-$0.15        &$-$0.05  & $-$0.17 &   0.32  &   0.21  &   0.09     \\
c23 &  2M18052388-2953056       &  $-$1.57 & $-$0.35 & +0.40 & 0.40 & +0.57          &$-$0.03  & $-$0.20 &   0.31  &   0.26  &   0.20     \\
c24 &  2M18142265-0904155       &  $-$0.85 & $-$0.20 & +0.20 & +0.33 & $-$0.32        & +0.29   & +0.18   &   0.37  &   0.26  &   0.21     \\
c25 &  2M17293482-2741164       &  $-$1.25 & $-$0.50 & +0.30 & +0.40 & +0.04          & +0.02   & $-$0.05 &  0.30   &  0.28   &  0.22      \\
c26 &  2M17341796-3905103       &  $-$0.89 & $-$0.40 & +0.25 & +0.40 & $-$0.28        & +0.35   & +0.35   &  0.34   &  0.36   &  0.23      \\
c27 &  2M17342067-3902066       &  $-$0.93 & $-$0.30 & +0.20 & +0.40 & +0.36$^{*}$     & +0.30   & +0.15   & 0.30   &  0.41   &  0.14      \\
c28 &  2M17503065-2313234       &  $-$0.88 & $-$0.10 & +0.00 & +0.35 & +0.09$^{*}$     & +0.23   & +0.28   & 0.36   &  0.22   &  0.18      \\
c29 &  2M18143710-2650147       &  $-$0.92 & $-$0.30 & +0.10 & +0.40 & $-$0.20        & +0.26   & +0.13   &  0.35   &  0.31   &  0.14      \\
c30 &  2M18150516-2708486       &  $-$0.83 & $-$0.05 & +0.00 & +0.31 & $-$0.04        & +0.24   & +0.14   &  0.37   &  0.22   &  0.15      \\
c31 &  2M18344461-2415140       &  $-$1.42 & $-$0.45 & +0.40 & +0.40 & $-$0.23        & +0.12   & $-$0.03 &  0.29   &  0.32   &  0.28      \\
\hline
\noalign{\smallskip}
\label{all}
\end{tabular}}

{\small NOTE: Na abundances in boldface are from the Value Added Catalogue (VAC) data derived with  BACCHUS Analysis of weak lines in APOGEE spectra (BAWLAS), and the symbol * means that for these stars, the Na abundance was considered reliable in \citet{barbuy23}.}

\end{table*}

\section{Conclusions}\label{sec7}

This is the third paper of a series dealing with the chemical abundance analysis of 58 stars selected to have [Fe/H]$<-$0.8 and orbits compatible with being members of a spheroidal component of the Galactic bulge.
In the present work, we have recomputed the abundances of the iron-peak elements V, Cr, Mn, Co, Ni, and Cu in spectra observed by the APOGEE collaboration. 
We confirm which lines  are suitable for the determination of these abundances in the
range of metallicities of   $-$2.0$<$[Fe/H ]$<-$0.8.

The abundances were compared with chemical-evolution models, which appear to reproduce the [X/Fe] versus [Fe/H] behaviour well in most cases.
In summary, 
V, Cr, and Ni tend to vary in lockstep with iron, as does  Co, although the present data on Co
 show a trend towards under-abundance for [Fe/H]$<$$-$1, and might be showing a secondary behaviour
 similar to that of Cu.
Mn  and Cu decrease with decreasing metallicity, and Cu appears to drop faster than the models predict for [Fe/H]$<$$-$0.8.
Finally, using abundance discriminators, together with kinematical and dynamical criteria,
our sample of 58 stars has the characteristics of an in situ sample.

\begin{acknowledgements}
B.B. and A.C.S.F. acknowledge grants from FAPESP, Conselho Nacional de Desenvolvimento Cient\'ifico e Tecnol\'ogico (CNPq) and Coordena\c{c}\~ao de Aperfei\c{c}oamento de Pessoal de N\'ivel Superior (CAPES) - Financial code 001. 
PS acknowledges Funda\c{c}\~ao de Amparo \`a Pesquisa do Estado de S\~ao Paulo (FAPESP) post-doctoral fellowships 2020/13239-5 and 2022/14382-1.
SOS acknowledges a FAPESP PhD fellowship no. 2018/22044-3.
A.P.-V., B.B., and S.O.S. acknowledge the DGAPA-PAPIIT grant IA103224.
PS, BB, HE, and SOS are part of the Brazilian Participation Group (BPG) in the Sloan Digital Sky Survey (SDSS), from the
Laborat\'orio Interinstitucional de e-Astronomia – LIneA, Brazil.
J.G.F-T gratefully acknowledges the grant support provided by Proyecto Fondecyt Iniciaci\'on No. 11220340, 
Proyecto Fondecyt Postdoc No. 3230001 (Sponsored by J.G.F-T)  and from the Joint Committee ESO-Government of Chile 2021 (ORP 023/2021), and 2023 (ORP 062/2023). 
F.A. acknowledges partial supported by the Spanish MICIN/AEI/10.13039/501100011033 and by "ERDF A way of making Europe" by the “European Union” through grant PID2021-122842OB-C21, and the Institute of Cosmos Sciences University of Barcelona (ICCUB, Unidad de Excelencia ’Mar\'{\i}a de Maeztu’) through grant CEX2019-000918-M. FA acknowledges the grant RYC2021-031683-I funded by MCIN/AEI/10.13039/501100011033 and by the European Union NextGenerationEU/PRTR.
D.M. gratefully acknowledges support from the Center for Astrophysics and Associated Technologies (CATA) by ANID BASAL projects ACE210002 and FB210003, and Fondecyt Project No. 1220724.
D.G. gratefully acknowledges the support provided by Fondecyt regular n. 1220264.
D.G. also acknowledges financial support from the Direcci\'on de Investigaci\'on y Desarrollo de
la Universidad de La Serena through the Programa de Incentivo a la Investigaci\'on de
Acad\'emicos (PIA-DIDULS).
The work of V.V.S. and V.M.P. is supported by NOIRLab, which is managed by the Association of Universities for Research in Astronomy (AURA) under a cooperative agreement with the U.S. National Science Foundation.
T.C.B. acknowledges support from grant PHY 14-30152; Physics Frontier Center/JINA Center for the Evolution of the Elements (JINA-CEE), and from OISE-1927130: The International Research  Network for Nuclear Astrophysics (IReNA), awarded by the US National Science Foundation.
Apogee project: Funding for the Sloan Digital Sky Survey IV has been provided by the Alfred P. Sloan Foundation, the U.S. Department of Energy Office of Science, and the Participating Institutions. SDSS acknowledges support and resources from the Center for High-Performance Computing at the University of Utah. The SDSS web site is www.sdss.org. 
SDSS is managed by the Astrophysical Research Consortium for the Participating Institutions of the SDSS Collaboration including the Brazilian Participation Group, the Carnegie Institution for Science, Carnegie Mellon University, Center for Astrophysics | Harvard \& Smithsonian (CfA), the Chilean Participation Group, the French Participation Group, Instituto de Astrof\'isica de Canarias, The Johns Hopkins University, Kavli Institute for the Physics and Mathematics of the Universe (IPMU) / University of Tokyo, the Korean Participation Group, Lawrence Berkeley National Laboratory, Leibniz Institut f\"ur Astrophysik Potsdam (AIP), Max-Planck-Institut f\"ur Astronomie (MPIA Heidelberg), Max-Planck-Institut f\"ur Astrophysik (MPA Garching), Max-Planck-Institut f\"ur Extraterrestrische Physik (MPE), National Astronomical Observatories of China, New Mexico State University, New York University, University of Notre Dame, Observat\'orio Nacional / MCTI, The Ohio State University, Pennsylvania State University, Shanghai Astronomical Observatory, United Kingdom Participation Group, Universidad Nacional Aut\'onoma de M\'exico, University of Arizona, University of Colorado Boulder, University of Oxford, University of Portsmouth, University of Utah, University of Virginia, University of Washington, University of Wisconsin, Vanderbilt University, and Yale University.
\end{acknowledgements}



\bibliographystyle{aa} 
\bibliography{bibliog6558}



\begin{appendix}

\section{Present and APOGEE-ASPCAP abundance results}

In Table \ref{results} are given the APOGEE uncalibrated stellar parameters, and the present resulting abundances for the
elements V, Cr, Mn, Co, Ni, and Cu.
In Table \ref{apogee} we report the abundances of V, Cr, Mn, Co, and Ni from APOGEE DR17 for the 58 sample stars. Copper was not measured by the APOGEE ASPCAP procedure.
The  values in bold face correspond to data from the Value Added Catalog (VAC), 
and analysed in BAWLAS (see text).

\begin{table*}[!ht]
\centering
\caption{Stellar parameters and abundances of V, Cr, Mn, Co, Ni, and Cu derived.} 
\resizebox{0.8\textwidth}{!}{
\begin{tabular}{cccccccccccccccccccccccccc}
\noalign{\smallskip}
\hline
\noalign{\smallskip}
\hbox{ID} & \hbox{ID}&
T$_{\rm eff}$ &\hbox{log~g} & \hbox{[Fe/H]} &  \hbox{v$_t$} &  [V/Fe] &  [Cr/Fe] & [Mn/Fe] & [Co/Fe] & [Ni/Fe] & [Cu/Fe] \\
\hbox{(internal)} & \hbox{(2MASS)} & \hbox{(K)} & &  &  \hbox{(km/s)} & & & & & &  \\  
 \noalign{\smallskip}
\noalign{\hrule}
\hline                   
b1 &2M17153858-2759467 & 3922.7 &  0.34 &  $-$1.65 & 2.62  & -0.05   &  +0.07 & $-$0.35 & $-$0.05 &+0.03 & +0.45  \\
b2 &2M17173693-2806495 & 3908.9 &  0.95 &  $-$0.97 & 2.20  & +0.05   &  +0.08 & $-$0.35 & +0.10 &+0.00 & $-$0.60  \\
b3 &2M17250290-2800385 & 3796.6 &  0.91 &  $-$0.82 & 2.39  & +0.05   &  +0.13 & $-$0.15 & +0.17 &-0.04 & +0.10 \\
b4 &2M17265563-2813558 & 4096.2 &  1.00 &  $-$1.32 & 1.89  & +0.15   &  +0.15 & $-$0.45 & +0.00 &+0.10 & $-$0.18 \\
b5 &2M17281191-2831393 & 4029.1 &  0.96 &  $-$1.19 & 1.73  & -0.15   &  $-$0.40 & -0.43 & $-$0.20 &-0.07 & $-$0.08  \\
b6 &2M17295481-2051262 & 4205.9 &  1.50 &  $-$0.85 & 1.71  &  ---    &  +0.15:& $-$0.45 & +0.00 &+0.03 & +0.25  \\
b7 &2M17303581-2354453 & 3863.0 &  0.77 &  $-$0.98 & 2.13  & +0.05   &  +0.20 & $-$0.20 & +0.15 &-0.02 & $-$0.10  \\
b8 &2M17324257-2301417 & 3668.2 &  0.79 &  $-$0.82 & 2.30  & +0.12   &  +0.13 & $-$0.18 & +0.15 &-0.02 & $-$0.35  \\
b9 &2M17330695-2302130 & 3566.6 &  0.35 &  $-$0.93 & 2.42  & +0.00   &  +0.15 & $-$0.17 & +0.16 &+0.00 & $-$0.20 \\
b10 &2M17344841-4540171& 3869.2 &  0.85 &  $-$0.88 & 2.16  & +0.05   &  +0.10 & $-$0.27 & +0.00 &+0.00 & $-$0.10   \\
b11 &2M17351981-1948329& 3553.5 &  0.44 &  $-$1.11 & 3.06  & -0.07   &  +0.08 & $-$0.25 & +0.00 &+0.02 & $-$0.27  \\
b12 &2M17354093-1716200& 3895.5 &  1.01 &  $-$0.87 & 2.02  & +0.10   &  +0.23 & $-$0.15 & +0.19 &-0.02 & $-$0.17   \\
b14 &2M17392719-2310311& 3643.3 &  0.67 &  $-$0.87 & 2.55  & +0.15   &  +0.20 & $-$0.12 & +0.14 &+0.07 & $-$0.45    \\
b15 &2M17473299-2258254& 4018.3 &  0.47 &  $-$1.74 & 2.12  & +0.00   &  ----  & $-$0.42 &$-$0.10 &+0.06  & $-$0.45   \\
b16 &2M17482995-2305299& 4213.6 &  1.24 &  $-$1.03 & 2.10  & ---     &  $-$0.15 & $-$0.50 & $-$0.12 &+0.00 & +0.00   \\
b17 &2M17483633-2242483& 3651.5 &  0.44 &  $-$1.09 & 2.58  & -0.02   &  +0.13 & $-$0.35 & $-$0.08 &+0.00 & $-$0.30  \\
b18 &2M17503263-3654102& 3893.5 &  0.64 &  $-$0.99 & 2.19  & +0.08   &  +0.12   & -0.30 & +0.03 &-0.05 & $-$0.18  \\
b19 &2M17552744-3228019& 4018.9 &  1.0  &  $-$1.06 & 2.00  & -0.05   &  $-$0.10 & $-$0.40 & $-$0.15 &-0.02 & +0.10  \\
b20 &2M18020063-1814495& 3988.8 &  0.80 &  $-$1.38 & 2.04  & -0.05   &   ---  & $-$0.30 & $-$0.09 &+0.05 & $-$0.40  \\
b21 &2M18050452-3249149& 3940.8 &  0.77 &  $-$1.16 & 2.08  & -0.10   &  +0.15 & $-$0.30 &$\geq$-0.05 &+0.04 & +0.08  \\
b22 &2M18050663-3005419& 3439.9 &  0.23 &  $-$0.92 & 2.52  & -0.15   &  +0.16 & $-$0.10 & +0.10 &-0.02 & $-$0.27  \\
b23 &2M18065321-2524392& 3893.1 &  0.95 &  $-$0.89 & 2.02  & +0.00   &  +0.10 & $-$0.23 & +0.06 &+0.00 & $-$0.45  \\
b24 &2M18104496-2719514& 4153.1 &  1.33 &  $-$0.82 & 2.05  &  ---    &   ---  & $-$0.25 & +0.03 &+0.09 & $-$0.20  \\
b25 &2M18125718-2732215& 3617.2 &  0.44 &  $-$1.31 & 2.64  &  ---    &  +0.05 & $-$0.15 & $-$0.30 &-0.02 & $-$0.07  \\
b26 &2M18200365-3224168& 3976.6 &  0.95 &  $-$0.86 & 1.94  & +0.00   &  +0.15 & $-$0.30 & +0.00 &+0.00 & $-$0.26  \\
b27 &2M18500307-1427291& 4076.0 &  1.23 &  $-$0.95 & 1.73  & -0.12   &  +0.15 & $-$0.40 & +0.00 &-0.03 & $-$0.53  \\
c1 &2M17173248-2518529 & 3977.0 &  1.0  &  $-$0.91 & 1.81  & +0.00   &  +0.20 & $-$0.35 & +0.00 &-0.02 & +0.05  \\
c2 &2M17285088-2855427 & 3838.0 &  0.63 &  $-$1.23 & 2.18  & -0.50   &  +0.15 & $-$0.40 & $-$0.15 &+0.00 & $-$0.50  \\
c3 &2M17301495-2337002 & 3814.0 &  0.69 &  $-$1.06 & 2.22  & -0.06   &  +0.15 & $-$0.43 & +0.00 &-0.06 & $-$0.35  \\
c4 &2M17453659-2309130 & 4133.1 &  1.27 &  $-$1.20 & 1.08  & ---     &  $-$0.35 & $-$0.38 &  ---  & ---  & $-$0.26  \\
c5 &2M17532599-2053304 & 3896.9 &  0.91 &  $-$0.87 & 2.10  & +0.08   &  +0.20 & $-$0.33 & +0.15 &+0.01 & $-$0.15 \\
c6 &2M18044663-3132174 & 3832.6 &  0.92 &  $-$0.90 & 2.22  & +0.07   &  +0.25 & $-$0.27 & +0.16 &+0.10 & +0.00   \\
c7 &2M18080306-3125381 & 4310.0 &  1.57 &  $-$0.90 & 1.48  & ---   &  +0.10 & $-$0.40 & +0.00:&-0.18 & $-$0.27   \\
c8 &2M18195859-1912513 & 4102.0 &  1.05 &  $-$1.24 & 1.78  & +0.12   &  +0.05 & $-$0.39 & $-$0.05 &+0.01 & +0.00   \\
c9 &2M17190320-2857321 & 4139.6 &  1.19 &  $-$1.20 & 1.83  &  ---    &  +0.00 & $-$0.38 & +0.00: & --- & $-$0.50  \\
c10 &2M17224443-2343053& 4058.3 &  1.02 &  $-$0.88 & 1.97  & +0.02   &  +0.15 & $-$0.24 & +0.00 &+0.00 & $-$1.00  \\
c11 &2M17292082-2126433& 3983.4 &  0.78 &  $-$1.27 & 2.59  & +0.25   &  +0.40 & $-$0.26 & +0.25 &+0.10 & +0.00  \\
c12 &2M17323787-2023013& 3865.7 &  1.03 &  $-$0.85 & 1.94  & +0.02   &  +0.20 & $-$0.25 & +0.03 &+0.02 & +0.00  \\
c13 &2M17330730-2407378& 4042.5 &  0.25 &  $-$1.90 & 1.88  &  ---    &  ---   & $-$0.54 & +0.00:&+0.00 & +0.15  \\
c14 &2M18023156-2834451& 3617.4 &  0.42 &  $-$1.19 & 3.02  & +0.08   &  $-$0.05 & $-$0.27 &+0.05 &-0.06 & +0.00   \\
c15 &2M17291778-2602468& 3844.3 &  0.71 &  $-$0.99 & 2.10  & $-$0.09   &  +0.18 & $-$0.31 & +0.03 &+0.02 & $-$0.30  \\
c16 &2M17310874-2956542& 4175.7 &  1.20 &  $-$0.93 & 2.07  &  ---    &  +0.08 & $-$0.37 & $-$0.06 &-0.04 & $-$0.05    \\
c17 &2M17382504-2424163& 3880.4 &  0.99 &  $-$1.05 & 1.55  & +0.00   &  +0.08 & $-$0.19 & +0.02 &+0.10 & $-$1.00  \\
c18 &2M17511568-3249403& 3921.2 &  0.98 &  $-$0.90 & 2.04  & +0.03   &  +0.17 & $-$0.42 & +0.06 &+0.01 & $-$0.15  \\
c19 &2M17552681-3334272& 4051.0 &  1.08 &  $-$0.89 & 1.98  & +0.30   &  +0.30 & $-$0.23 & +0.06 &+0.01 & +0.00  \\
c20 &2M18005152-2916576& 4158.9 &  1.04 &  $-$1.02 & 2.21  & +0.13   &   ---  & $-$0.32 & +0.10 &-0.04 & +0.00  \\
c21 &2M18010424-3126158& 3773.1 &  0.68 &  $-$0.83 & 2.20  & +0.05   &  +0.15 & $-$0.29 & +0.00 &+0.00 & $-$0.10  \\
c22 &2M18042687-2928348& 4164.7 &  0.88 &  $-$1.21 & 2.14  & ---   &  +0.00:& $-$0.54 & $-$0.15:&+0.00 & +0.00   \\
c23 &2M18052388-2953056& 4252.9 &  0.92 &  $-$1.57 & 1.92  &  ---    &  +0.17:& $-$0.59 &  ---  &+0.00 & $-$0.30   \\
c24 &2M18142265-0904155& 3920.5 &  1.12 &  $-$0.85 & 2.13  &  ---    &  +0.17:& $-$0.32 & +0.10 &+0.02 & $-$0.45:   \\
c25 &2M17293482-2741164& 4143.5 &  1.03 &  $-$1.25 & 1.85  &  ---    &  +0.18:& $-$0.45 & +0.06 &-0.03 & $-$0.80:  \\
c26 &2M17341796-3905103& 4163.5 &  1.42 &  $-$0.89 & 1.84  &  ---    &  +0.25 & $-$0.37 & +0.00:&+0.01 & $-$0.10  \\
c27 &2M17342067-3902066& 4380.4 &  1.40 &  $-$0.93 & 1.99  &  ---    &  ---   & $-$0.40 & +0.00:&+0.01 & ---  \\
c28 &2M17503065-2313234& 3819.4 &  0.98 &  $-$0.88 & 2.10  & +0.00   &  +0.10 & $-$0.24 & +0.14 &+0.06 & +0.00  \\
c29 &2M18143710-2650147& 4240.5 &  1.30 &  $-$0.92 & 1.97  &  ---   &  +0.30 & $-$0.31 & +0.05 &+0.00 & $-$0.30   \\
c30 &2M18150516-2708486& 3833.4 &  1.0  &  $-$0.83 & 2.14  & +0.01   &  +0.10 & $-$0.23 & $-$0.03 &+0.00 & +0.00   \\
c31 &2M18344461-2415140& 4294.5 &  1.09 &  $-$1.42 & 1.83  &  ---   &  ---   & $-$0.50 & +0.00 &+0.00 & +0.30   \\
\hline
\noalign{\smallskip}
\hline 
\label{results}
\end{tabular}}
\begin{minipage}{13cm}
\vspace{0.1cm}
\small The internal identification of stars with numbers
starting with b or c, corresponds to spectra from DR14 and DR16 respectively.
\end{minipage}
\end{table*}

\begin{table*}
\centering
\caption{Abundances of V, Cr, Mn, Co, and Ni from APOGEE DR17 for the 58 sample stars.
}
\resizebox{0.65\textwidth}{!}{
\begin{tabular}{llrrrrr}
\hline
\noalign{\smallskip}
ID & ID & [V/Fe] & [Cr/Fe] & [Mn/Fe] & [Co/Fe] & [Ni/Fe] \\ 
\noalign{\smallskip}
\hline
b1& 2M17153858-2759467  & ---     &$-$0.278 &$-$0.375 &$-$0.053 & 0.025  \\
b2& 2M17173693-2806495  &{\bf 0.201} & 0.028   &$-$0.310 & 0.022   &$-$0.006  \\
b3& 2M17250290-2800385  &{\bf 0.169}& 0.030     & ---     & 0.170   & 0.042  \\
b4& 2M17265563-2813558  & 0.141 & 0.282     &$-$0.391 &$-$0.131 & 0.068  \\
b5& 2M17281191-2831393  &$-$0.059 &$-$0.644 &$-$0.239 &$-$0.200 &$-$0.065   \\
b6& 2M17295481-2051262  &$-$0.201 &$-$0.214 &$-$0.358 & 0.047   & 0.029   \\
b7& 2M17303581-2354453  &$-$0.069 & 0.021   &$-$0.319 & 0.059   &$-$0.019   \\
b8& 2M17324257-2301417  &{\bf 0.232}& ---     & ---     & 0.147   & 0.080  \\
b9& 2M17330695-2302130  &$-$0.403 & ---     & ---     & 0.160   & 0.033    \\
b10& 2M17344841-4540171 &{\bf 0.256} & 0.004   &$-$0.266 & 0.073   & 0.039  \\
b11& 2M17351981-1948329 & 0.108   & ---     & ---     & ---     & 0.068 \\
b12& 2M17354093-1716200 &$-$0.005 &$-$0.023 &$-$0.255 & 0.143   & 0.062  \\
b13& 2M17390801-2331379 &{\bf 0.210} & 0.022   & ---   & 0.192     & 0.083  \\
b14& 2M17392719-2310311 & 0.018   & ---     & ---   & 0.142     & 0.006  \\
b15& 2M17473299-2258254 &$-$0.155 & ---     &$-$0.470 &$-$0.252 &$-$0.060  \\
b16& 2M17482995-2305299 &$-$0.480 &$-$0.013 &$-$0.438 &$-$0.026 &$-$0.028  \\
b17& 2M17483633-2242483 &{\bf 0.191}& ---     & ---     & 0.088   & 0.020   \\
b18& 2M17503263-3654102 &$-$0.016 &$-$0.065 &$-$0.275 & 0.033   & 0.042  \\
b19& 2M17552744-3228019 &{\bf 0.347}&$-$0.132 &$-$0.314 & 0.022   & 0.004 \\
b20& 2M18020063-1814495 &$-$0.831 &$-$0.349 &$-$0.346 &$-$0.155 &$-$0.037  \\
b21& 2M18050452-3249149 &$-$0.081 & 0.015   &$-$0.297 &$-$0.050 & 0.030  \\
b22& 2M18050663-3005419 & ---     & ---     & ---     & ---     & ---    \\
b23& 2M18065321-2524392 &$-$0.134 & 0.016   &$-$0.280 &$-$0.019 & 0.004  \\
b24& 2M18104496-2719514 &$-$0.027 &$-$0.115 &$-$0.250 & 0.005   & 0.008  \\
b25& 2M18125718-2732215 &$-$0.350 & ---     & ---     &$-$0.012 & 0.029  \\
b26& 2M18200365-3224168 &{\bf 0.060}& 0.026   &$-$0.296 & 0.057   & 0.000  \\
b27& 2M18500307-1427291 &$-$0.459 &$-$0.174 &$-$0.403 &$-$0.026 &$-$0.041  \\
c1& 2M17173248-2518529  &$-$0.221 &$-$0.099 &$-$0.299 & 0.010   &$-$0.064  \\
c2& 2M17285088-2855427  &$-$0.814 &$-$0.092 &$-$0.401 &$-$0.067 &$-$0.070  \\
c3& 2M17301495-2337002  &{\bf 0.211}& 0.032   &$-$0.304 & 0.068   & 0.020  \\
c4& 2M17453659-2309130  &$-$0.381 &$-$0.820 &$-$0.417 & ---     &$-$0.115  \\
c5& 2M17532599-2053304  &{\bf 0.164}&$-$0.083 &$-$0.256 & 0.084   & 0.052  \\
c6& 2M18044663-3132174  & 0.036 & 0.000     &$-$0.221 & 0.157   & 0.056  \\
c7& 2M18080306-3125381  &$-$0.460 &$-$0.117 &$-$0.397 &$-$0.425 &$-$0.082  \\
c8& 2M18195859-1912513  & 0.101 &$-$0.060   &$-$0.390 &$-$0.303 &$-$0.058  \\
c9& 2M17190320-2857321  &$-$0.628 &$-$0.359 &$-$0.384 & 0.109   &$-$0.023  \\
c10& 2M17224443-2343053 &{\bf 0.080}& 0.004   &$-$0.237 & 0.000   & 0.033\\
c11& 2M17292082-2126433 &$-$0.120 & 0.041   &$-$0.307 & 0.063   &$-$0.037  \\
c12& 2M17323787-2023013 &{\bf 0.091}& 0.006   &$-$0.261 & 0.031   &$-$0.002  \\
c13& 2M17330730-2407378 & 0.030   & ---     &$-$0.542 &$-$0.402 &$-$0.056  \\
c14& 2M18023156-2834451 & 0.152   & ---     &  ---    & 0.196   & 0.141 \\
c15& 2M17291778-2602468 &$-$0.130 &$-$0.005 &$-$0.312 &$-$0.017 & 0.007  \\
c16& 2M17310874-2956542 &$-$0.440 &$-$0.222 &$-$0.338 &$-$0.060 & 0.039  \\
c17& 2M17382504-2424163 &$-$0.123 &$-$0.259 &$-$0.194 & 0.023   & 0.163   \\
c18& 2M17511568-3249403 & 0.010   & 0.119   &$-$0.332 & 0.064   &$-$0.013   \\
c19& 2M17552681-3334272 &$-$0.013 & 0.073   &$-$0.274 & 0.057   & 0.024  \\
c20& 2M18005152-2916576 & 0.130   &$-$0.009 &$-$0.275 & 0.101   &$-$0.021  \\
c21& 2M18010424-3126158 &{\bf 0.141} & 0.045     & ---     & 0.074   & 0.022  \\
c22& 2M18042687-2928348 &$-$0.321 &$-$0.591 &$-$0.537 &$-$0.248 &$-$0.044  \\
c23& 2M18052388-2953056 &$-$0.881 & 0.168   &$-$0.590 & 0.000   & 0.007  \\
c24& 2M18142265-0904155 &$-$0.182 &$-$0.224 &$-$0.284 & 0.009   & 0.062  \\
c25& 2M17293482-2741164 & ---     &$-$0.088 &$-$0.555 & 0.056   &$-$0.084  \\
c26& 2M17341796-3905103 &$-$1.052 &$-$0.146 &$-$0.368 &$-$0.498 & 0.073  \\
c27& 2M17342067-3902066 &$-$0.134 &$-$0.390 &$-$0.496 &$-$0.085 & 0.041  \\
c28& 2M17503065-2313234 &$-$0.009 & 0.066   &$-$0.244 & 0.138   & 0.046  \\
c29& 2M18143710-2650147 & 0.053   &$-$0.026 &$-$0.356 &$-$0.031 &$-$0.010 \\
c30& 2M18150516-2708486 &$-$0.004 &$-$0.017 &$-$0.226 & 0.061   & 0.054  \\
c31& 2M18344461-2415140 & 0.156   &$-$0.128 &$-$0.654 &$-$0.218 &$-$0.0905\\ \hline \hline
\label{apogee}
\end{tabular}}
\begin{minipage}{13cm}
\vspace{0.1cm}
\small Values in bold face are from VAC-BAWLAS.
\end{minipage}
\end{table*}

\section{Non-Local thermodynamic Equilibrium corrections}

 Table \ref{nlte} provides the NLTE corrections for the lines available
in \citet{bergemann10}, 
for the elements Cr, Mn, and Co. We only adopted the final
values for the Co abundances, given that for Cr and Mn there are no 
corrections for all the lines studied, and for Cr and Mn the corrections
appear too high.

\begin{table*}
\centering
\caption{NLTE corrections (in dex) for the Cr~I, Mn~I, Co~I lines available in nlte.mpia.de website. }
\resizebox{0.8\textwidth}{!}{
\begin{tabular}{lcccccc}
\hline
\hbox{ID} & \hbox{CrI} & \hbox{CrI} & \hbox{CrI} & \hbox{MnI} & \hbox{MnI}  & \hbox{CoI} \\
   & 15680.063 & 15860.214 (GAP) & 16015.323 (BLEND)&  15217.793&  15262.702 & 16757.719\\
     \hline
  b1      &    0.134   &   0.124  &    0.114  &   -0.008  &    0.147  &   0.009 \\
  b2      &    0.139   &   0.131  &    0.122  &    0.091  &    0.758  &  -0.046 \\
  b3      &    0.113   &   0.104  &    0.097  &    0.105  &    0.840  &  -0.059 \\
  b4      &    0.215   &   0.205  &    0.192  &    0.049  &    0.431  &  -0.010 \\
  b5      &    0.185   &   0.176  &    0.166  &    0.051  &    0.543  &  -0.024 \\
  b6      &    0.204   &   0.193  &    0.184  &    0.213  &   no-conv &  -0.039 \\
  b7      &    0.127   &   0.119  &    0.111  &    0.074  &    0.688  &  -0.046 \\
  b8      &    0.090   &   0.083  &    0.077  &    0.079  &    0.695  &  -0.065 \\
  b9      &    0.075   &   0.069  &    0.065  &    0.042  &    0.521  &  -0.062 \\
 b10      &    0.126   &   0.118  &    0.110  &    0.100  &    0.818  &  -0.054 \\
 b11      &    0.078   &   0.073  &    0.069  &    0.021  &    0.365  &  -0.043 \\
 b12      &    0.133   &   0.125  &    0.116  &    0.124  &    0.935  &  -0.053 \\
 b14      &    0.086   &   0.080  &    0.075  &    0.061  &    0.592  &  -0.063 \\
 b15      &    0.163   &   0.152  &    0.140  &    0.019  &    0.162  &   0.018 \\
 b16      &    0.225   &   0.215  &    0.204  &    0.102  &    0.797  &  -0.025 \\
 b17      &    0.090   &   0.085  &    0.081  &    0.018  &    0.402  &  -0.041 \\
 b18      &    0.128   &   0.121  &    0.113  &    0.062  &    0.657  &  -0.042 \\
 b19      &    0.170   &   0.161  &    0.152  &    0.087  &    0.711  &  -0.034 \\
 b20      &    0.177   &   0.170  &    0.159  &   -0.001  &    0.313  &  -0.014 \\
 b21      &    0.153   &   0.146  &    0.138  &    0.036  &    0.531  &  -0.028 \\
 b22      &   weak     &  weak    &   weak    &    0.075  &    0.470  &  weak   \\
 b23      &    0.135   &   0.126  &    0.117  &    0.113  &    0.888  &  -0.053 \\
 b24      &    0.181   &   0.172  &    0.165  &    0.199  &   no-conv &  -0.041 \\
 b25      &    0.093   &   0.087  &    0.081  &    0.010  &    0.289  &  -0.029 \\
 b26      &    0.145   &   0.137  &    0.128  &    0.142  &    1.008  &  -0.051 \\
 b27      &    0.178   &   0.168  &    0.159  &    0.150  &    1.012  &  -0.041 \\
 c1       &    0.151   &   0.142  &    0.133  &    0.134  &    0.989  &  -0.049 \\
 c2       &    0.133   &   0.126  &    0.119  &    0.028  &    0.470  &  -0.028 \\
 c3       &    0.122   &   0.115  &    0.107  &    0.055  &    0.605  &  -0.043 \\
 c4       &    0.253   &   0.239  &    0.224  &    0.110  &    0.722  &  -0.026 \\
 c5       &    0.132   &   0.124  &    0.115  &    0.114  &    0.898  &  -0.053 \\
 c6       &    0.125   &   0.116  &    0.107  &    0.092  &    0.767  &  -0.054 \\
 c7       &    0.256   &   0.241  &    0.228  &    0.193  &   no-conv &  -0.028 \\
 c8       &    0.215   &   0.205  &    0.193  &    0.071  &    0.535  &  -0.016 \\
 c9       &    0.241   &   0.230  &    0.215  &    0.089  &    0.603  &  -0.020 \\
c10       &    0.165   &   0.156  &    0.147  &    0.149  &    1.018  &  -0.044 \\
c11       &    0.167   &   0.160  &    0.151  &    0.009  &    0.365  &  -0.017 \\
c12       &    0.129   &   0.120  &    0.111  &    0.128  &    0.959  &  -0.056 \\
c13       &    0.170   &   0.161  &    0.152  &    0.098  &    0.192  &   0.043 \\
c14       &    0.088   &   0.083  &    0.078  &    0.012  &    0.313  &  -0.033 \\
c15       &    0.123   &   0.115  &    0.107  &    0.068  &    0.670  &  -0.046 \\
c16       &    0.196   &   0.187  &    0.180  &    0.134  &    0.987  &  -0.034 \\
c17       &    0.149   &   0.139  &    0.128  &    0.097  &    0.872  &  -0.045 \\
c18       &    0.141   &   0.132  &    0.123  &    0.117  &    0.900  &  -0.050 \\
c19       &    0.163   &   0.154  &    0.146  &    0.148  &    1.014  &  -0.043 \\
c20       &    0.199   &   0.190  &    0.182  &    0.087  &    0.761  &  -0.026 \\
c21       &    0.106   &   0.099  &    0.092  &    0.083  &    0.720  &  -0.062 \\
c22       &    0.230   &   0.219  &    0.204  &    0.054  &    0.489  &  -0.001 \\
c23       &    0.302   &   0.284  &    0.265  &    0.088  &    0.317  &   0.042 \\
c24       &    0.139   &   0.130  &    0.120  &    0.137  &    0.980  &  -0.050 \\
c25       &    0.226   &   0.216  &    0.203  &    0.071  &    0.510  &  -0.009 \\
c26       &    0.195   &   0.185  &    0.177  &    0.182  &   no-conv &  -0.038 \\
c27       &    0.283   &   0.270  &    0.255  &    0.138  &    0.982  &  -0.010 \\
c28       &    0.126   &   0.116  &    0.107  &    0.104  &    0.832  &  -0.056 \\
c29       &    0.218   &   0.207  &    0.197  &    0.145  &    1.015  &  -0.033 \\
c30       &    0.127   &   0.117  &    0.107  &    0.121  &    0.913  &  -0.058 \\
c31       &    0.328   &   0.306  &    0.285  &    0.097  &    0.414  &   0.034 \\
\hline   
\hline         
\label{nlte}
\end{tabular}}
\begin{minipage}{13cm}
\vspace{0.1cm}
\small The keywords no-conv, non-LTE did not converge and weak: The line is too weak (EW $<$ 1mA).
\end{minipage}
\end{table*}

\end{appendix}

\end{document}